\documentclass[prd,preprintnumbers,superscriptaddress,showpacs,eqsecnum,twocolumn]{revtex4}

\usepackage{graphicx}
\usepackage{amsmath}
\usepackage{amssymb}

\def\be{\begin{equation}}
\def\ee{\end{equation}}
\def\ba{\begin{eqnarray}}
\def\ea{\end{eqnarray}}
\def\F{\mathcal{F}}
\def\th{\textrm{\mbox{\tiny{th}}}}
\def\Tobs{T_{\textrm{\mbox{\tiny{obs}}}}}
\def\Tcoh{T_{\textrm{\mbox{\tiny{coh}}}}}
\def\nd{{\mathbf{n}}_d}
\def\SSB{\textrm{\mbox{\tiny{ssb}}}}
\def\ns{\textrm{\mbox{\tiny{NS}}}}
\def\min{\textrm{\mbox{\tiny{min}}}}
\def\max{\textrm{\mbox{\tiny{max}}}}

\begin{document}

\newcommand*{\AG}{Max-Planck-Institut f\"ur
    Gravitationsphysik, Albert-Einstein-Institut, Am M\"uhlenberg 1,
    D-14476 Golm, Germany}\affiliation{\AG}
\newcommand*{\BB}{Departament de F\'{\i}sica, Universitat de les Illes
  Balears, Cra. Valldemossa Km. 7.5, E-07122 Palma de Mallorca,
  Spain}\affiliation{\BB}
\newcommand*{\CU}{Cardiff University, Cardiff, CF2 3YB, United 
 Kingdom}\affiliation{\CU}
\newcommand*{\ROF}{Dip. di Fisica, Universit\`a di Roma ``La
    Sapienza'' P.le A. Moro 2, I-00185 Rome, 
    Italy}\affiliation{\ROF}
\newcommand*{\ROP}{INFN Sezione di Roma, P.le A. Moro 2, I-00185 Rome,
    Italy}\affiliation{\ROP}
 
\title{The Hough transform search for continuous gravitational waves}

\author{Badri Krishnan}\email{badri.krishnan@aei.mpg.de}\affiliation{\AG}
\author{Alicia M. Sintes}\email{alicia.sintes@uib.es}\affiliation{\AG}\affiliation{\BB}
\author{Maria Alessandra Papa}\email{maria.alessandra.papa@aei.mpg.de}\affiliation{\AG} 
\author{Bernard F. Schutz}\email{schutz@aei.mpg.de}\affiliation{\AG}\affiliation{\CU}
\author{Sergio Frasca}\email{sergio.frasca@roma1.infn.it}\affiliation{\ROF}\affiliation{\ROP}
\author{Cristiano Palomba}\email{cristiano.palomba@roma1.infn.it}\affiliation{\ROP}

    
  \begin{abstract}

    This paper describes an incoherent method to search for
    continuous gravitational waves based on the \emph{Hough transform}, a
    well known technique used for detecting patterns in digital
    images.  We apply the Hough transform to detect patterns in the
    time-frequency plane of the data produced by an earth-based
    gravitational wave detector.  Two different flavors of searches
    will be considered, depending on the type of input to the Hough
    transform: either Fourier transforms of the detector data or the
    output of a coherent matched-filtering type search. We present the
    technical details for implementing the Hough transform algorithm
    for both kinds of searches, their statistical properties, and
    their sensitivities.  

  \end{abstract}

  \pacs{04.80.Nn, 04.30.Db, 95.55.Ym, 07.05.Kf, 97.60.Gb}
  \preprint{AEI-2004-050}
  \maketitle


\section{Introduction}
\label{sec:intro}

Rapidly rotating neutron stars are expected to be the primary sources of
continuous gravitational waves, and the current generation of earth-based
gravitational wave detectors might be able to detect them.  Recent
analysis of data from  the first science runs of the
LIGO~\cite{S1:detector,ligo1, ligo2} and
GEO~\cite{S1:detector,GEO1,GEO2} interferometric detectors has already 
led to upper limits on the gravitational waves emitted by the pulsar
J1939+2134 and its equatorial ellipticity~\cite{S1:pulsar}. The
analysis of future science runs is expected to lead to upper limits
below other astrophysical constraints, and eventually to detections.    

The analyses presented in~\cite{S1:pulsar} were based on the coherent
integration of the detectors' output for the entire observation time
(approximately 17 days) and used a Bayesian time-domain method and a
frequentist frequency-domain~\cite{jks} approach. The searches were
not computationally expensive, targeting a single known pulsar and
processing only a narrow frequency band of about 0.5 Hz around the
pulsar emission frequency for a fixed sky location and spin-down rate
known from radio observations. 

Future continuous wave searches will involve searching longer data
stretches (of order weeks to months) for unknown sources over a large
frequency band, vast portions of the sky and spin-down parameter
values.  It is well known that the computational cost of coherent
techniques for searches of this type is absolutely
prohibitive~\cite{bccs}. Thus hierarchical methods have been
proposed. 

In hierarchical strategies incoherent techniques (less sensitive and
less computationally expensive) are used to scan the data and the
parameter space for interesting candidates which are then followed up
with coherent searches. Different strategies can be envisaged that
combine the data incoherently. All methods use, in some way,
the power from the Fourier transforms of short stretches of data: in
the frequency bins where the signal is present there should
systematically be an excess of power. In order to compensate for the
frequency modulation imposed on the signal by the Earth's motion and
the pulsar's spin-down during the observation 
period, one must use not the power from the same frequency bins in each
successive Fourier transform, but rather from the bins where one expects the
signal peak to be. 

In the so called stack-slide method, one ``slides''
the frequency bins of each Fourier transform to line-up the signal
peaks and then simply sums the power~\cite{bc}. The Hough transform
method can be seen 
as a variation on this where, after the sliding, one sums not 
the power but just zeros and ones, depending on whether the power in the
frequency bin exceeds a threshold or meets some other criterion.
Whereas in low signal-to-noise conditions in Gaussian noise, the
standard power summing method is possibly optimal, the Hough transform
method might be more robust in the presence of large spectral
disturbances.  
To see this, consider the case when a large spectral disturbance is present
only in a single Fourier transform. This could
have a very large effect on the power sum statistic, but no matter how
large, this spectral line could only add $+1$ to the Hough statistic.

The Hough transform is a robust
parameter estimator of multi-dimensional patterns in images and it finds
many applications in astronomical data analysis \cite{storkey,ballester,ca}.
In the context of
image processing, it provides robustness against missing data points
or discontinuous features \cite{ik}.  It was initially
developed by Paul Hough to analyze bubble chamber pictures at CERN,
and later patented by IBM~\cite{hough1,hough2}.
It is currently being used to analyze data from the LIGO and GEO
detectors. The codes employed for these analyses are 
freely available as part of the LIGO Algorithms Library~\cite{lal}. The
VIRGO project~\cite{virgo,virgo97} is also setting up a similar
hierarchical search pipeline.  Studies of hierarchical strategies can
be found in~\cite{fp,sp,pss,bfpr,afp,f}.

This paper is organized as follows: section \ref{sec:prelim} briefly
describes the expected waveforms from an isolated spinning neutron star and
summarizes the general strategy of a hierarchical search. Section
\ref{sec:houghbasics} presents the general idea of the Hough 
transform and section \ref{sec:houghnondemod} describes its
implementation for non-demodulated input data, and section
\ref{sec:statistics} studies its statistical properties.  Section
\ref{sec:demod} describes the Hough search using demodulated input
data and finally section \ref{sec:conclusion} summarizes our main
results.

\section{Preliminaries}
\label{sec:prelim}

\subsection{The signal from a pulsar}
\label{subsec:pulsar}

In this subsection we fix our notation and briefly review the expected 
gravitational wave signal from a spinning neutron star.  Further
details about the pulsar
signal can be found in~\cite{jks}; a concise review of the possible 
physical mechanisms that may be causing pulsars to emit gravitational
waves can be found in~\cite{S1:pulsar}.  For our purposes, we only
need the form of the gravitational wave signal as seen by an Earth
based detector.  

\begin{figure}
  \begin{center}
  \includegraphics[height=5cm]{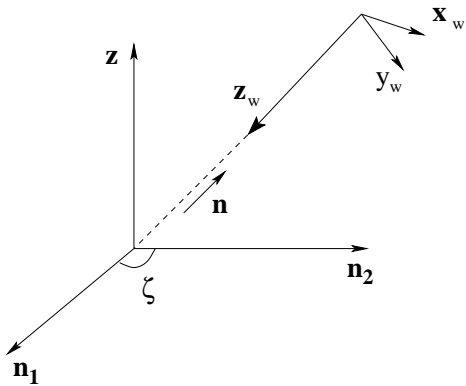}
  \caption{ The detector frame and the wave frame. }\label{fig:waveframe}
  \end{center}
\end{figure}

Let $\mathbf{n}_1$ and $\mathbf{n}_2$ denote the unit vectors pointing
along the arms of the detector and denote by $\zeta$ the angle between 
the arms.  Let $\mathbf{z}$ be the unit vector parallel to
$\mathbf{n}_1\times\mathbf{n}_2$.   Apart from the detector frame
$(\mathbf{n}_1,\mathbf{n}_2,\mathbf{z})$, we also have the wave frame
$(\mathbf{x}_w,\mathbf{y}_w,\mathbf{z}_w)$ in which the unit vector
$\mathbf{z}_w$ is along the direction of propagation of the wave and
$(\mathbf{x}_w,\mathbf{y}_w,\mathbf{z}_w)$ form a right-handed
orthonormal system.  Finally, $\mathbf{n}=-\mathbf{z}_w$ is the unit vector pointing
in the direction of the neutron star; see figure \ref{fig:waveframe}.  
The spacetime metric $g_{\mu\nu}$ can be written as a perturbation of
the flat metric $\eta_{\mu\nu}$: $g_{\mu\nu} = \eta_{\mu\nu} +
h_{\mu\nu}$.  The received gravitational wave $h_{\mu\nu}$ has the form 
\be \label{eq:perturbation}
h_{\mu\nu}(t) = h_+(t)\left(\mathbf{e}_{+}\right)_{\,\mu\nu} +
h_\times(t)\left(\mathbf{e}_{\times}\right)_{\,\mu\nu} 
\ee
where $\mathbf{e}_+ = \mathbf{x}_w\otimes\mathbf{x}_w
-\mathbf{y}_w\otimes\mathbf{y}_w$ and $\mathbf{e}_\times =
\mathbf{x}_w\otimes\mathbf{y}_w+\mathbf{y}_w\otimes\mathbf{x}_w$, and
$t$ denotes clock time at the location of the (moving, accelerating) detector, 
which we refer to as {\em detector time}. The waveforms for the two
polarizations are
\be \label{eq:pluscross} 
h_+(t) = A_+\cos\Phi(t) \, , \qquad  h_\times(t) = A_\times\sin\Phi(t) 
\ee
where $\Phi(t)$ is the phase of the gravitational wave and
$A_{+,\times}$ are the amplitudes; $A_{+,\times}$ are
constant in time and
depend on the other pulsar parameters such as its
rotational frequency, moments of inertia, the orientation of its
rotation axis, its distance from Earth etc.  The phase $\Phi(t)$ takes
its simplest form when the time coordinate used is $t_\ns$, the proper
time in the rest frame of the neutron star:  
\be\label{eq:phasemodel} \Phi_\ns(t_\ns) = \phi_0 + 2\pi \sum_{n=0}^{s}
\frac{f_{(n)}^{(\ns)}}{(n+1)!} t_\ns^{n+1} \ee
where $\phi_0$, $f_{(0)}^{(\ns)}$ and $f_{(n)}^{(\ns)}$ ($n\geq 1$) are
respectively the phase, instantaneous
frequency and the spin-down parameters in the rest frame of the star
at the fiducial start time $t_\ns=0$, and $s$ is the number of
spin-down parameters included in our search.  

We refer the reader to~\cite{jks} for the expression of $\Phi(t)$ in
the detector frame as a function of detector time.  For our purposes, we only need to know
that the instantaneous frequency $f(t)$ of the wave as observed by the detector 
is given, to a very good approximation, by the familiar
non-relativistic Doppler formula:
\be\label{eq:master1}
f(t) - \hat{f}(t) = \hat{f}(t)\frac{ {\bf v} (t)\cdot\bf{n}}{c} 
\ee
where $\mathbf{v}(t)$ is the detector velocity in the Solar System
Barycenter (SSB) frame and $\hat{f}$ is given by
\be \label{eq:f0hat}
\hat{f}(t) = f_{(0)} +
\sum_{n=1}^{s}\frac{f_{(n)}}{n!}\left(t -t_0 +
\frac{\Delta\mathbf{r}(t)\cdot\mathbf{n}}{c}\right)^n  
\ee
where $t_0$ is the fiducial detector time at the start 
of the observation, the $f_{(n)}$ are the spin-down parameters as
measured in the SSB frame (these need not be equal to the
$f_{(n)}^{(\ns)}$; see~\cite{jks}), and
$\Delta\mathbf{r}(t):=\mathbf{r}(t) -\mathbf{r}(t_0)$ with $\mathbf{r}(t)$
being the position of the detector in the SSB frame at time $t$.       
We have also assumed the neutron star to be moving with uniform 
speed relative to the Sun and is so far away that there are no 
observable proper-motion effects. (These could be taken into account 
if necessary, at the cost of introducing further parameters.)  

The detector output is a linear combination of $h_+$ and
$h_\times$:
\be\label{eq:waveform} h(t) = F_+(\mathbf{n},\psi)\,h_+(t) +
F_\times(\mathbf{n},\psi)\,h_\times(t) \ee 
where  $F_{+,\times}$ are known as the the antenna pattern functions
of the detector and depend on the direction $\mathbf{n}$ to the star
and also on the polarization angle $\psi$ which determines the
orientation of the $(\mathbf{x}_w,\mathbf{y}_w)$ axes in their plane.  In addition,
the antenna pattern functions also depend on the detector parameters
such as its latitude, the angle $\zeta$ between its arms, and the
azimuth of the bisector of the arms.  Due to the motion of the Earth,
$F_{+,\times}(\mathbf{n},\psi)$ depend implicitly on time and for
notational convenience, we shall usually denote the antenna pattern
functions as $F_{+,\times}(t)$. Thus, the received signal is 
both amplitude- and frequency-modulated.  

The search method described in this paper depends on finding a 
signal whose frequency evolution fits the pattern produced by the
Doppler shift and the spin-down.
The parameters which determine this pattern are the ones which appear
in equation (\ref{eq:master1}), namely, $(f_{(0)},
\{f_{(n)}\},{\mathbf{n}})$; these parameters will be collectively
denoted by $\vec{\xi}$.

The amplitudes $A_{+,\times}$ are determined by the other pulsar
parameters such as the orientation of its axis, its ellipticity, its distance
from Earth etc.  The search method presented in this paper depends only
on the phase model of equation (\ref{eq:phasemodel}).
The exact form of the amplitudes is model dependent.  As an
illustrative example, consider the wave emitted by a deformed spinning
neutron star as in~\cite{S1:pulsar}. If $f_r$ is the rotational
frequency of the star, the frequency of the gravitational
wave is $2f_r$.  The additional parameters determining this
component of the pulsar signal are $\iota$ and $h_0$ where
$\iota$ is the angle between the pulsar's axis of rotation and
the vector $\mathbf{z}_w=-\mathbf{n}$, and $h_0$ 
characterizes the amplitude of the emitted gravitational 
wave. The amplitudes $A_{+,\times}$ are:
\ba \label{eq:amplitude2}
A_{+} &=& \frac{1}{2}h_0 (1+\cos^2\iota) \,,\\
A_{\times} &=& h_0\cos\iota\, .
\ea
If we assume the emission mechanism is due to deviations of the pulsar's shape
from perfect axial symmetry, then the
amplitude $h_0$ will be
\be \label{eq:h0} h_0 = \frac{16\pi^2G}{c^4}\frac{I_{zz}\epsilon
f_r^2}{d} \ee
where $d$ is the distance of the star from Earth, $I_{zz}$ is the
z-z component of the star's moment of inertia with the $z$-axis being
its spin axis, and $\epsilon:= (I_{xx}-I_{yy})/I_{zz}$ is the
equatorial ellipticity of the star.  Among all the quantities
appearing in this equation, the value of $\epsilon$ is by far the
most uncertain.  Typical values are expected to be $\sim 
10^{-8}$ for standard neutron stars and values of $\sim 10^{-6}$ are expected to be
the maximum values~\cite{bccs}.  There is also a very small
uncertainty in the value of $f_r$ because the pulsar could have a
(presently unobservable) radial velocity. This would produce a Doppler
shift between the true value of $f_r$ in the neutron star (NS) frame,
and its measured 
value on Earth.   Assuming typical values of the radial velocity to be the
same as the typically measured transverse velocities ($\sim 500$
km/s, see e.~g. \cite{ktr}), we get an uncertainty in $f_r$ of $\sim 0.1$\%.

\subsection{A multi-stage hierarchical search}
\label{sec:hierarchical}

Consider performing a blind search for pulsars using a bank of
templates and relying only on coherent matched filter techniques.  
Since a larger observation
time implies better resolution in the space of frequency, spin-downs
and sky-positions, the number of templates increases rapidly as a
function of the total observation time.   A typical example is an
all-sky search for young, fast pulsars, i.e. for hypothetical signals
with frequency $\hat{f}< f_\max = 1000$Hz and spin-down ages greater
than $\tau > \tau_\min = 40$yr.  Let $s$
be the number of spin-down parameters that we search over and let
$\Tobs$ be the total observation time.  The number of templates
required for this search has been calculated in equation (6.3) of
\cite{bccs}:  
\be
N_p \approx {\textrm{max}}_{s\in\{0,1\ldots\}}[N_sF_s(\Tobs)]
\ee
where
\be
N_s =
\left(\frac{f_\max}{1\textrm{kHz}}\right)^{s+2}\left(\frac{40\textrm{yr}}{\tau_\min}\right)^{s(s+1)/2} 
\ee
gives the spin-down scaling and $F_s$ is a function that depends on the 
observation time; for large observation times, $F_s\propto \Tobs^5$. 
We have taken the maximum allowed fractional mismatch in observed
signal power between the signal and the template to be 0.3. 
For example, if $s=2$, assuming the observation time is significantly
longer than a day, equation (6.7) of~\cite{bccs} approximates to :
\be
F_2(\Tobs) \approx 2.2\times 10^7 \times \left(\frac{\Tobs}{1\textrm{day}}\right)^5 \,.
\ee
Thus, even for a $10$ day search over two spin-down parameters, $N_p
\approx 2\times 10^{12}$.  The computational requirements for a
search over these many templates is also estimated in~\cite{bccs}. It
turns out that for the $10$ day long search, if we wished to analyze the
data in roughly real time, we would require a computational power of
$\sim 10^{8}$ GFlops;  for reference, the fastest supercomputers
ca.~2004 can do `only' $\sim 10^4$ GFlops.  Even if we 
insisted on searching over only a single spin-down parameter, for an
observation period of only $10$ days, the
computational requirement turns out to be $\sim 10^{5}$ GFlops.  
\emph{We therefore conclude that a search over any significant portion of
parameter space for unknown pulsars is not possible in the foreseeable
future if we restrict ourselves to fully coherent methods. }

\begin{figure}
  \begin{center}
  \includegraphics[height=7cm]{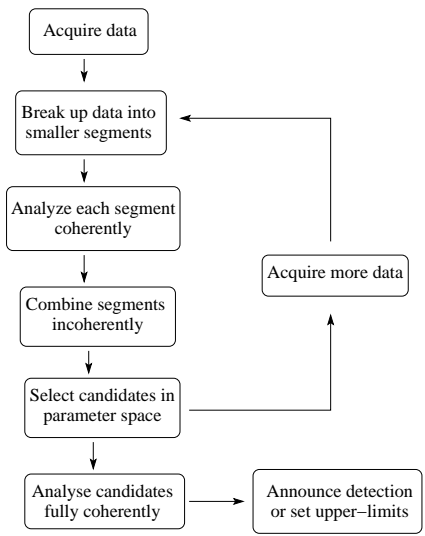}
  \caption{A hierarchical scheme for the analysis of large parameter space
  volumes for continuous wave searches.  Each step only analyzes the
  regions in parameter space that have not been discarded by any of
  the previous steps.}\label{fig:hierarchical} 
  \end{center}
\end{figure}

One possible way to perform such a blind search would be to exploit
the fact that $N_p$ increases faster than 
linearly with $\Tobs$. Thus if we break up the data set into smaller
segments, it might be feasible to analyze each data segment
coherently.  An incoherent method is then used as a computationally
inexpensive and sub-optimal way of combining the outputs of the
different coherent segments. This would be one step in a multi-stage
hierarchical scheme; see figure \ref{fig:hierarchical}.   

In this scheme, we start with a
data stream covering a total observation time $\Tobs$. Divide the
available data into smaller segments and analyze each segment 
coherently. The results of this coherent analysis of the different 
segments are combined incoherently.  The output
of the incoherent step is a set of possible pulsar candidates.   
If necessary, acquire fresh data and repeat the above
procedure analyzing only the candidates selected
by the previous step. Once this procedure has been iterated the desired
number of times and the number of candidates in parameter space is
small enough, the candidates are analyzed by using the entire data stream
coherently.  The final output of the search is, of course,
either a detection or an upper limit.

The exact number of times the
incoherent step must be repeated and the thresholds that one must set at
each stage are decided by optimizing the sensitivity subject to
the obvious constraints on the desired signal strength we wish to
detect, the desired confidence level and the amount of total data available.
Preliminary investigations of this optimization are reported in
\cite{bc} and more detailed results will be presented elsewhere~\cite{cgk}.

Hierarchical searches like this are typically effective only when 
looking for signals that, in the final coherent search over the whole data 
set, have relatively high signal-to-noise ratio. The method only works  
if the incoherent step succeeds in reducing the number of points 
in parameter space that one must search over.  A signal that is only, say,  
at two-sigma in the final step will be too weak in the initial shorter 
coherent transforms to be selected by any criterion that would eliminate 
other (\lq\lq pure noise") parameter points.  
Remarkably, this does not actually reduce the sensitivity of a hierarchical 
search by much over the hypothetical fully coherent search we
described above. The   
reason is that the size of the parameter space is so large that, even 
in a fully coherent search, signals must be unusually strong in order to 
be detected with enough significance to be recognized. In our case, a 
fully coherent signal-to-noise ratio of 10 or more is needed for a significant
detection over a period of several months, and we will see in 
equation~(\ref{eq:sensitivity}) below and the subsequent discussion that
our incoherent methods do worse than this by factors of between 2 
and 5, while permitting much larger regions of parameter space to 
be surveyed.

Finally we mention one important detail, namely, the nature of the
coherent analysis of each data segment.  In this paper we consider two
possible 
alternatives. The first alternative is just to use the Fourier
transform of data segments that are so short that no frequency modulation 
or spin-down is measurable. These transforms are called SFTs 
(Short time base-line Fourier Transform) and may represent up to 
30 minutes of data.  The candidates for the incoherent step are
selected based on the \emph{normalized} SFT power, i.e. on the power
divided by the noise floor estimate.  

If longer coherent stages are required for better sensitivity,
then one must use \emph{demodulated} data, i.e.\ remove the effects of 
Earth's spin and orbital motion and also of the pulsar spin-down.  This 
demodulation must be done separately for different regions of the sky
and spin-down parameter space, but it also brings in other parameters, 
such as the polarization angle $\psi$, because of the effects of 
amplitude modulation. These extra parameters, which are not part of 
our Hough-transform search space, can be eliminated by requiring the 
coherent stage to produce the $\F$-statistic described in~\cite{jks} and used
in~\cite{S1:pulsar} for analyzing the data from the first science runs
of the LIGO and GEO detectors.  In this case, we would select
frequency bins based on the value of the $\F$-statistic.  The search
based on SFTs will be called the \emph{non-demodulated} search and is
described in sections \ref{sec:houghnondemod} and
\ref{sec:statistics}.  The search using the $\F$-statistic is the
search with \emph{demodulated} data and is described in section
\ref{sec:demod}.

\section{The Hough transform}
\label{sec:houghbasics}

As mentioned in the introduction, the Hough transform is a robust
parameter estimator for patterns in digital images.  
It can be used to identify the parameter(s) of a curve which best fits a
set of given  points.  In the last two decades, the Hough transform
has become a standard tool in the domain of artificial vision for the 
recognition of patterns that can be parameterized like straight lines,
polynomials, circles, etc.

For our purposes,
a pattern is a collection $\mathcal{C}$ of smooth 
hypersurfaces \footnote{The word hypersurface refers to a sub-manifold
of unit co-dimension.  The generalization to surfaces of higher
co-dimension is straightforward but we shall not discuss it in this
paper.} in some differentiable manifold $M$.  
Assume that there is a manifold $\Sigma$ of parameters which
describes elements of $\mathcal{C}$; i.e. there exists a function
$f:\Sigma\rightarrow\mathcal{C}$ providing a one-one association
between points in $\Sigma$ and elements of $\mathcal{C}$.

A simple example is the case when $M$ is $\mathbb{R}^2$ with
coordinates $x$  
and $y$, and $\mathcal{C}$ is the collection of straight lines in this 
$(x,y)$ plane.  Since all straight lines are described by an equation
of the form $y=mx+c$ (the master equation), the parameter space $\Sigma$ is also
$\mathbb{R}^2$, with coordinates $(m,c)$ -- the slope and the
$y$-intercept of the straight lines.  The function $f$ maps the
point $(m,c)$ to the straight line $y=mx+c$.  
The relevant example for our purposes is the case when the manifold
$\Sigma$ represents the pulsar parameters $\vec{\xi} =
(f_{(0)},\{f_{(n)}\},\mathbf{n})$ and $M$ is the
time-frequency plane. The pattern in $M$ is described by
the Doppler shift formula of equation (\ref{eq:master1}).  Each value
of $\vec{\xi}$ determines the frequency evolution $f(t)$ and thus
determines a curve in the time-frequency plane.  

\begin{figure}
  \begin{center}
  \includegraphics[height=6cm]{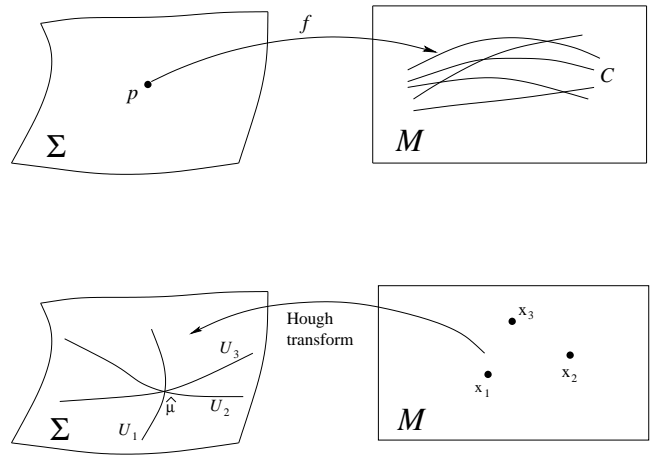}
  \caption{A schematic depiction of the Hough transform in the absence
  of noise.  The top figure shows the parameter space $\Sigma$ and the
  space of observations $M$.  The space of expected patterns in 
  is a set $\mathcal{C}$ of hypersurfaces in $M$. The
  function $f:\Sigma\rightarrow\mathcal{C}$ provides a one-one
  correspondence between $\Sigma$ and $\mathcal{C}$.  The lower figure
  shows the Hough transform itself: Every observation $x_i$ is mapped
  via the Hough transform into a hypersurface $\mathcal{U}_i$ in
  parameter space which is consistent with the observation.  The
  intersection of all the $\mathcal{U}_i$'s contains the true source
  parameter $\hat{\mu}$.}\label{fig:houghbasic} 
  \end{center}
\end{figure}

Given a set of observations $\{x_i\}$ with each $x_i$ belonging to
$M$, we ask if 
there is an underlying pattern describing these points and whether
this pattern is described by a hypersurface belonging to
$\mathcal{C}$.  Consider first the idealized case when there is no
noise and the points $\{x_i\}$ actually do
follow the pattern and lie on one single hypersurface belonging to
$\mathcal{C}$ corresponding to the parameter value
$\hat{\mu}\in\Sigma$.   
How would we go about finding $\hat{\mu}$ if we were
given the collection $\{x_i\}$?  For every $x_i$, the idea is to
first find the set of points $\mathcal{U}_i$ in parameter space
consistent with $x_i$; the true parameter value $\hat{\mu}$ must
certainly lie within this set.  In the straight line example, all the
lines passing through the observed point would be consistent with that
observation.  Repeating this for every observation $x_i$, we obtain a
collection of subsets $\{\mathcal{U}_i\}$. The true parameter value
$\hat{\mu}$ must lie in each $\mathcal{U}_i$ and therefore it must
also lie in the intersection
\be \hat{\mu} \in \bigcap_{i} \mathcal{U}_i \, .\ee
See figure \ref{fig:houghbasic}.  If $k$ is the dimensionality of
$\Sigma$, then we need at least $k$ 
different $x_i$'s in order to ensure that $\hat{\mu}$ can be found uniquely.  
Thus, in this idealized noiseless case, we would need
only two observations to detect a straight line.  Similarly for the
pulsar case, equation (\ref{eq:master1}) is the master equation
and if we were searching for $s$ spin-down parameters, we would need
only $3+s$ observations to determine the pulsar parameters.  This is,
of course, not true when noise is present.  

In realistic situations, the presence of noise will ensure that, in
general, there is no point which is consistent with \emph{all} the
$x_i$'s, in other words, $\cap_i\,\mathcal{U}_i$ is the
empty set.  In this case we proceed as follows:  to each
$\mu\in\Sigma$, assign an integer $n(\mu)$ (the \emph{number
count}), which is equal to the number of 
$\mathcal{U}_i$'s which contain $\mu$.  The result is then a
histogram in parameter space.  This procedure, which maps a set
of observations to a histogram in parameter space, will be called the
Hough transform.  The best candidate for the true
parameter $\hat{\mu}$ is then the point at which the number count is
maximal.  Alternatively, we could set an appropriate threshold $n_\th$ on the
number count and select all points in $\Sigma$ at which the number
count exceeds $n_\th$.  These selected parameter space points would be
candidates for a possible detection and, if we were performing a
multi-stage hierarchical search, would be further analyzed in the next
step.   

\begin{figure}
  \begin{center}
  \includegraphics[height=2.6cm]{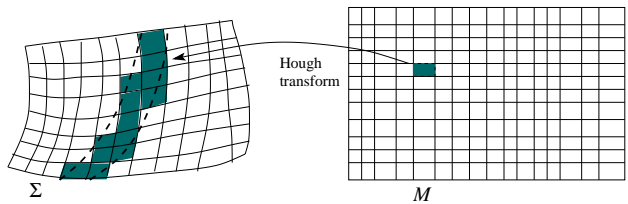}
  \caption{A schematic view of the Hough transform for the discrete
  case. An observation consists of a pixel in $M$ which goes
  over to the region enclosed between the dotted lines under the Hough
  transform.  This in turn leads to a selection of pixels in parameter
  space.  The shaded pixels are the ones which get selected and are
  the ones consistent with the observation. }\label{fig:houghdiscrete}
  \end{center}
\end{figure}

In real experiments, we cannot perform a parameter space search
with infinite resolution.  Therefore we need to consider the discrete case
when we have a finite resolution for the observations and also a grid
on parameter space.  In this case, observations correspond to
\emph{pixels} in $M$. The general procedure is essentially the same as in the
discrete case and is depicted schematically in figure
\ref{fig:houghdiscrete}: we look for pixels in parameter space which
are consistent with the
observations.  There is, however, one technical difference namely,
since each observation is an extended region in $M$,
the points in parameter space consistent with this observation do not
constitute a sharp hypersurface $\mathcal{U}_i$. Each pixel instead
gives a \emph{region} $\tilde{\mathcal{U}}_i$ bounded by two such 
hypersurfaces. Given such a region, we can then select pixels in
parameter space.  Since a pixel in
parameter space might intersect more than one $\tilde{\mathcal{U}}_i$,
we need an unambiguous criterion to select pixels in parameter space in
order to ensure that each pixel gets selected at most once by an
observation.  Given such a criterion, we can continue the earlier
strategy and construct a 
histogram in parameter space by assigning a number count to each pixel
in parameter space.  The pixel with the largest number count is
our best candidate for a detection.  

\section{The Hough transform with non-demodulated data}
\label{sec:houghnondemod}
  
\begin{figure}
  \begin{center}
  \includegraphics[height=6cm]{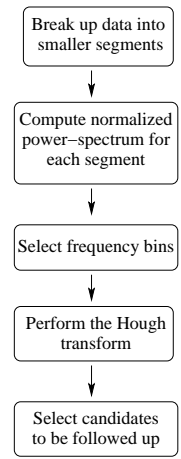}
  \caption{A single stage of a hierarchical continuous wave search involving the
  Hough transform.  The starting point is to break up the data with
  total observation time $\Tobs$ into $N$ segments and to compute the
  Fourier transform of each segment.  The next step is to select
  frequency bins from each SFT by setting a threshold on the
  normalized power spectrum and use the selected frequency bins to
  construct a Hough map.  The output is then a set of candidates in
  parameter space obtained by setting a threshold on the Hough number 
  count. }\label{fig:nondemod}
  \end{center}
\end{figure}

The steps involved in a single incoherent stage of the search are
outlined in figure \ref{fig:nondemod}.  In this search, one starts
by breaking up the input data of duration $\Tobs$ into $N$ 
segments each with a duration of
$\Tcoh$, which would be equal to $\Tobs/N$ if there were no gaps in
the data.  Except for precisely two exceptions, namely equations 
(\ref{eq:sensitivity}) and (\ref{eq:sensitivityF}), all the
equations in this paper will be valid even in the presence of gaps; we
shall \emph{not} assume $\Tcoh = \Tobs/N$.  This is important because
in practice, the real data stream will inevitably have gaps in it
representing times when the detector is not in lock or the data is not
reliable. 

The next step is to compute the Fourier
transform of each data segment to obtain $N$ SFTs.  Select
frequency bins in each SFT by setting a threshold on the normalized power
spectrum.  This produces a distribution of points in the time-frequency 
plane --- the manifold $M$ --- most of which are noise but some excess 
of which are hopefully present along one or more signal patterns given
by equation (\ref{eq:master1}). Having selected points in the time-frequency
plane, go through the Hough transform algorithm to obtain the Hough
map, i.e. the histogram, in parameter space $\Sigma$.  The details
follow.  

\subsection{Notation and conventions}
\label{subsec:notation}

We assume that the $N$ different data segments have the same time 
duration.  Label the 
different segments by $a=0,1\ldots (N-1)$ and denote the start time of
each segment by $t_a$ which will often be called the \emph{timestamp}
of the $a^{th}$ data segment. Let each segment consist of $M$ data
points.

Let us now focus on the $a^{th}$ data segment which covers the time
interval $[t_a,t_a+\Tcoh]$.  Let $x(t)$ be the detector output which is 
sampled at times $t_j = t_a+j\Delta t$ with $j=0,1,\ldots
(M-1)$. Here the data segment has been subdivided into $M$
sub-segments with the times $t_j$ defined to be at the start of each
sub-segment; this implies $\Delta t = \Tcoh/M$.
Denote the sequence of data points thus obtained by $\{x_j\}$ where 
$x_j\equiv x(t_j)$.  

Our convention for the Discrete Fourier Transform (DFT) of
$\{x_j\}$ is 
\be 
\tilde{x}_k = \Delta t \sum_{j=0}^{M-1}x_j e^{-2\pi ijk/M}
\ee
where $k=0,1\ldots (M-1)$.  
For $0\leq k \leq \lfloor M/2 \rfloor$, the frequency index $k$ corresponds to a
physical frequency of $f_k= k/\Tcoh$ with $\lfloor .\rfloor$ denoting
the integer part of a given real number.  The values 
$\lfloor M/2 \rfloor < k \leq M-1$ correspond to negative frequencies given
by $f_k = (k-M)/\Tcoh$. 

The detector output $x(t)$ at any time $t$ is the sum of
noise $n(t)$ and a possible gravitational wave signal $h(t)$ of known form:
\be x(t) = n(t)+h(t) \,.\ee
In the remainder of this paper, unless otherwise stated, the
stochastic process $n(t)$ is
assumed to be stationary and Gaussian with zero mean.  
   
In the continuous case, when the observation time is infinite, 
the single-sided power spectral density (PSD)
$S_n(f)$ for $f\ge 0$ is defined as the Fourier transform of the
auto-correlation function: 
\be
S_n(f) = 2\int_{-\infty}^\infty \langle n(t)n(0)\rangle e^{-2\pi
ift}dt 
\ee
where $\langle \cdot\rangle$ denotes the ensemble average.

The \emph{normalized power} is a dimensionless quantity defined as 
\be \label{eq:rhodef} \rho_k =
\frac{|\tilde{x}_k|^2}{\langle |\tilde{n}_k|^2\rangle}
\ee
It can be shown that $\langle |\tilde{n}_k|^2\rangle$ is related to the PSD:
\be
\langle |\tilde{n}_k|^2\rangle \approx \frac{M\Delta t}{2}S_n(f_k) =
\frac{\Tcoh}{2} S_n(f_k)\,.
\ee
Thus:
\be
\rho_k \approx \frac{2|\tilde{x}_k|^2}{\Tcoh S_n(f_k)} \,.
\ee
Naturally, the PSD must be estimated in a way that is not biased by
any signal power that may be present.

\subsection{Implementation}
\label{subsec:houghdetails}

The implementation choices we present here mostly correspond to those
that have been implemented in the Hough analysis code which is
publicly available as part of the LIGO Algorithms Library (LAL)
\cite{lal}, and will be used to analyze the data from the GEO and LIGO
detectors.\\

\noindent \textbf{\emph{Restriction on $\Tcoh$:}} For non-demodulated data,
the coherent integration time $\Tcoh$, i.e. the time-baseline
of the SFTs, cannot be arbitrarily large.  This restriction comes
about because we would like the signal power to be concentrated
in half a frequency bin but the signal frequency is changing in time due
to the Doppler modulation and also due to the spin-down of the star.  
If $\dot{f}$ is the time-derivative of the
signal frequency at any given time, in order for the signal not to shift by more
than half a frequency bin, we must have $|\dot{f}|\Tcoh < 
(2\Tcoh)^{-1}$,
i.e.
\be
 \Tcoh < \sqrt{\frac{1}{2|\dot{f}|_{\max}}} 
\ee
where by $|\dot{f}|_{\max}$ we mean the maximum possible value
of $|\dot{f}|$ for all allowed values of the shape parameters
$\vec{\xi}$.  
The time variation of $f(t)$ is given by equation (\ref{eq:master1})
and is due to two effects: the spin-down of
the star, and the Doppler modulation due to the Earth's motion.  We
shall assume that the Doppler modulation is the dominant
effect \footnote{This approximation implies an upper bound on the value of
the spin-down parameters that we can search over; this will be
discussed later in this subsection in greater detail.}.  Thus we
can estimate $\dot{f}$ by keeping $\hat{f}$ fixed and differentiating
${\bf{v}}(t)$ in equation (\ref{eq:master1}): 
\be
\dot{f} \approx \frac{\hat{f}}{c}\frac{d{\bf{v}}}{dt}\cdot {\bf{n}}
\leq \frac{\hat{f}}{c}\left|\frac{d{\bf v}}{dt}\right|
\, .
\ee
The important contribution to the acceleration $d{\bf{v}}/dt$ is from
the daily rotation of the Earth:
\be |\dot{f}|_{\max} =  \frac{\hat{f}}{c}\cdot \frac{v_e^2}{R_e} =
\frac{\hat{f}}{c}\cdot\frac{4\pi^2R_e}{T_e^2} \ee
where $v_e$ is the magnitude of the velocity of Earth around its axis,
$T_e$ the length of a day and $R_e$ the radius of Earth. Substituting
numerical values we get
\be \label{eq:tcoh} \Tcoh < 50 \,\textrm{min} \times
\sqrt{\frac{500\,\textrm{Hz}}{\hat{f}}} \,.\ee
In this paper, we shall mostly use $\Tcoh=30$min as the canonical
reference value.\\

\noindent \textbf{\emph{Selecting frequency bins:}}  The simplest
method of selecting frequency bins is to set a threshold $\rho_\th$ on
$\rho_k$; i.e. we select the $k^{th}$ frequency bin if $\rho_k \geq
\rho_\th$ and reject it otherwise.  Alternatively~\cite{f,allen02}, we could impose
additional conditions such as requiring that $\rho_k > \rho_{k+1}$ and
$\rho_k > \rho_{k-1}$, i.e. the $k^{th}$ bin is selected if $\rho_k$
exceeds the threshold and is, in addition, a 
\emph{local maxima}. This can be extended further by including more
than just the two neighboring frequency bins.  While it is relatively
easy to investigate these alternate strategies for non-demodulated
data, the analysis becomes more complicated for demodulated data.
Furthermore, while these alternate methods might be more robust
against spectral disturbances, the analysis of the statistics follows
the same general scheme and the results are not qualitatively
different.  Thus, for the purposes of this paper, we will describe
only the simple thresholding scheme for selecting frequency bins.
The optimal choice of the threshold $\rho_{\th}$ is described below 
in subsection \ref{subsec:optimalthr}.\\

\noindent \textbf{\emph{Solving the master equation:}}  As discussed in the
previous section, to perform the Hough transform, we must find
all the points in parameter space which are consistent with a given 
observation.  In this case, the observation is a frequency $f_k$
selected using a threshold $\rho_\th$ in say, the $a^{th}$ SFT
corresponding to a timestamp $t_a$.  This corresponds to a frequency bin
$(f_k-\frac{1}{2}\delta f,f_k+\frac{1}{2}\delta f)$ where $\delta f =
\Tcoh^{-1}$ is the frequency resolution of the SFT. The parameters
$\vec{\xi}$ of the
signal are the frequency, spin-down parameters, and the sky-positions:
$\vec{\xi} = (f_{(0)},\{f_{(n)}\},{\bf{n}})$. Corresponding to
$(f_k-\frac{1}{2}\delta f,f_k+\frac{1}{2}\delta f)$, we must
find all the possible values of $\vec{\xi}$ which satisfy the
master equation (\ref{eq:master1}).  

\begin{figure}
  \begin{center}
  \includegraphics[height=5cm]{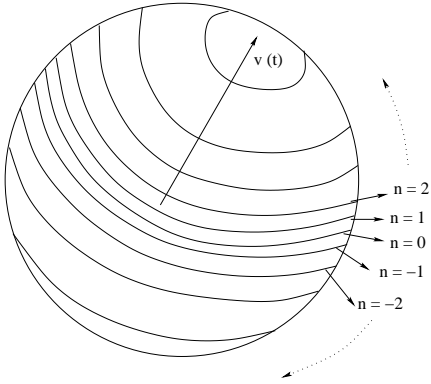}
  \caption{The set of sky positions consistent with a given frequency bin
  at a given time correspond to annuli on the celestial sphere.  These annuli are
  centered on the velocity vector $\bf{v}$, they are thin when
  perpendicular to $\bf{v}$ and thick when nearly parallel. The circle
  with with the label $n=0$ corresponds to $f=\hat{f}$. }\label{fig:annuli}
  \end{center}
\end{figure}

\begin{figure}
  \begin{center}
  \includegraphics[height=4cm]{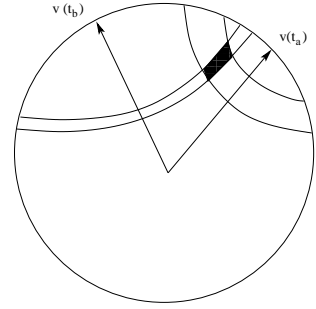}
  \caption{Two intersecting annuli.  The two timestamps $t_a$ and
  $t_b$ are sufficiently different from each other so that the
  velocities $\bar{\mathbf{v}}(t_a)$ and $\bar{\mathbf{v}}(t_b)$ are
  not parallel to each other.  This causes the annuli constructed at
  different timestamps to intersect.  The shaded region is the
  intersection and there is a corresponding region (not shown) on the
  far side of the sphere.}\label{fig:twoannuli}
  \end{center}
\end{figure}

To understand this better, let us
first fix the values of the frequency $f_{(0)}$ and the spin-down parameters
$\{f_{(n)}\}$ so that $\hat{f}(t)$ is also fixed.  Ignore, for the
moment, the frequency resolution $\delta f$.  From 
equation (\ref{eq:master1}), we see that all the values of 
${\bf{n}}$ consistent with the observation $f(t)$ must satisfy   
\be \label{eq:annulus1}
\cos\phi = \frac{ {\bf v}(t)\cdot {\bf{n}}}{v(t)} = 
\frac{c}{v(t)} \frac{f(t) - \hat{f}(t)}{\hat{f}(t)}
\ee
where $\phi$ is the angle between ${\bf v}(t)$ and ${\bf{n}}$.
This implies that the angle $\phi$ must be
constant; in other words, the set of sky positions consistent with an
observation $f(t)$ form a \emph{circle} in the celestial sphere centered on
the vector $\bf{v}$ (see figure \ref{fig:annuli}) \footnote{The
observation time is now not just a single instant $t$ but is instead a
time interval $(t_a,t_a+\Tcoh)$.  Therefore, instead of using the
detector velocity $\mathbf{v}(t)$, we must now use the \emph{average}
detector velocity $\bar{\mathbf{v}}(t_a)$.  This is relevant when $\Tcoh$
becomes comparable to a day, as it will be in the demodulated
case. }. 
If the frequency $f(t)$ is smeared over 
a frequency bin $(f_k-\frac{1}{2}\delta f,f_k+\frac{1}{2}\delta
f)$, the set of points consistent with an observation
must correspond to an \emph{annulus} the
width $\delta\phi$ of which is easily calculated using equation
(\ref{eq:annulus1}):
\be\label{eq:annulisize}
\delta\phi \approx \frac{c}{v}\frac{\delta f}{\hat{f}\sin\phi}\, .
\ee
The annuli are very thick at points where $\sin\phi$ is small,
i.e. when ${\bf{n}}$ is almost parallel or anti-parallel to
${\bf v}(t)$ and very thin when perpendicular.  This is depicted
schematically in figure \ref{fig:annuli}.  The circles on the
celestial sphere are labeled by an integer $n$ such that the
frequency $f=\hat{f} + n\delta f$ corresponds to the angle $\phi_n$
given by 
\be \label{eq:annulilabels} \cos\phi_n = \frac{nc\delta f}{v\hat{f}} \,.\ee
The lower limit on the
width of the annuli is provided by setting $\phi=\pi/2$ in equation
(\ref{eq:annulisize}): 
\ba &&\left(\delta\phi\right)_{\min} = \frac{c}{v}\frac{\delta
  f}{\hat{f}} = \frac{c}{v\hat{f}\Tcoh}  \\ && =  4.8\times
  10^{-3}\,\textrm{rad}\times \left(\frac{1\textrm{hr}}{\Tcoh}\right)
\left(\frac{500\,\textrm{Hz}}{\hat{f}}\right) \left(\frac{10^{-4}}{v/c}\right)\, .\nonumber
\label{eq:minannulus} \ea
The upper limit on the annuli width $(\delta\phi)_\max$ is
found by setting $\sin\phi \approx \phi \approx (\delta\phi)_{\max}$ which
gives
\ba
&& \left(\delta\phi\right)_{\max} =
\sqrt{\frac{\delta f}{\hat{f}}\frac{c}{v}} = \sqrt{\frac{c}{v\hat{f}\Tcoh}}
\\ && = 7.3\times
10^{-2}\,\textrm{rad} \times 
\left(\frac{1\textrm{hr}}{\Tcoh}\right)^{\frac{1}{2}} \left(\frac{500 
\textrm{Hz}}{\hat{f}}\right)^{\frac{1}{2}}\left(\frac{10^{-4}}{v/c}\right)^{\frac{1}{2}}
\,. \nonumber \label{eq:maxannulus}
\ea
Therefore, the thick annuli are about $10$ times thicker than the thin
ones.  
Different frequency bins selected at the same time will correspond to
non-intersecting annuli as shown in figure \ref{fig:annuli}.  
However, for frequency bins selected from SFTs at different time
stamps, say $t_a$ and $t_b$, the annuli will usually intersect because the
velocity vectors $\bar{\mathbf{v}}(t_a)$ and $\bar{\mathbf{v}}(t_b)$
will not, in general, be parallel to each other; see figure
\ref{fig:twoannuli}. 
p\\

\noindent \textbf{\emph{Resolution in the space of sky-positions:}} 
In order to search for
pulsar signals in a given portion of the sky, we must choose a tiling
for the sky patch.  Given the calculation of the annuli width above,
we choose the pixel size $\delta\theta$ of the grid to be some fraction, say
at most half, of the width $(\delta\phi)_{\min}$ of the thinnest
annulus. While this educated guess for the pixel size is sufficient 
for the purposes of this paper, the correct choice of pixel size in the sky
patch, and also in the entire parameter space, should use the
parameter space metric introduced in~\cite{owen}.  The analysis of
this metric for the Hough search will be presented elsewhere, and for
now we shall simply use $\delta\theta=\frac{1}{2}\,(\delta\phi)_{\min}$.  

Having selected an annulus and having chosen a tiling on our
sky-patch, we now need a criterion for selecting a pixel if it
intersects an annulus.  Our criterion is to select a pixel if its
center lies within an annulus.  Under such a criterion,
a given pixel can then be selected by at
most one annulus and the pixels selected by all the annuli together
will exactly cover the sphere.  \\

\noindent \textbf{\emph{Resolution in the space of spin-down parameters:}} 
In the absence of a proper analysis of the parameter
space metric, we shall just use the obvious estimate for the
resolution $\delta f_{(n)}$:
\be \label{eq:deltafk}
\delta f_{(n)} = n! \frac{\delta f}{\Tobs^n}\,.
\ee
As an example, for the first spin-down parameter:
\be \label{eq:deltaf1}
\delta f_{(1)} = (2.1\times 10^{-10}\textrm{Hz/s})\times
\frac{30\textrm{days}}{\Tobs}\cdot \frac{1800\textrm{s}}{\Tcoh} \,. 
\ee
We now need to choose the range of values $-f_{(n)}^\max < f_{(n)} <
f_{(n)}^\max$ and the largest number of spin-down parameters
$s_\max$ to be searched over. 
Assuming that the pulsar's frequency evolution is well-represented 
by a Taylor expansion, we get
\be \label{eq:fkmax}
f_{(n)}^\max = n! \frac{\hat{f}_\max}{\tau^n}
\ee
where $\tau$ is the age of the pulsar and $\hat{f}_\max$ is the
largest intrinsic frequency that we search over.  
We include the $n^{th}$ spin-down parameter in our search
only if the resolution defined by equation (\ref{eq:deltafk}) is not
too coarse compared to $f_{(n)}^\max$:
\be \label{eq:conditionsmax}
\delta f_{(n)} < f_{(n)}^{\textrm{\mbox{\tiny{max}}}} \,.
\ee
Since $\Tobs\ll\tau$, $f_{(n)}^\max$ decreases with increasing $n$
much faster than $\delta f_{(n)}$; this implies that there must exist
a value $s_\max$ such that equation (\ref{eq:conditionsmax}) is
satisfied for all $n\le s_\max$ and is violated for all
$n>s_\max$.  Any spin-down parameter of order greater
than $s_\max$ does not significantly affect the result of the Hough
transform.  As an example, if
we wish to search for pulsars whose age is at least $\tau = 40$yrs,
then for $\hat{f}_\max = 1000$Hz, it is easy to check that we get
$s_\max = 3$.  In other words, to look for pulsars which are as young
as $40$yr, we must include at least 3 spin-down parameters in our
search.   

On the other hand, in some cases, computational requirements might
dictate that we can only search over, say, one spin-down parameter.
This automatically sets a lower limit on the age of the pulsar that we
can search over because then the
second spin-down parameter must satisfy $\delta f_{(2)} >
f_{(2)}^\max$ which leads to
\be
\tau > 155\textrm{yr} \times \frac{\Tobs}{30\textrm{days}}\cdot
\left(\frac{\hat{f}_\max}{1000\textrm{Hz}}
\cdot\frac{\Tcoh}{1800\textrm{s}}\right)^{1/2}   
\,. 
\ee

Finally, the finite length of $\Tcoh$ itself leads to a
lower bound on $\tau$.  If $f_{(n)}$ is too large, then the signal
power may no longer be concentrated in a single frequency bin and the
assumption of neglecting spin-down parameters which was used to derive
equation (\ref{eq:tcoh}) will no longer be valid.  To be certain that 
the spin-down will not cause the signal to move by more than half a
frequency bin, we must have $f_{(n)}^\max\Tcoh^n < n!\delta f/2$ which 
implies
\be \label{eq:taunondemod}
\tau > \left(\frac{2\hat{f}_\max\Tcoh^{n+1}}{n!}\right)^{1/n} \,.
\ee
The most stringent limit is obtained for $n=1$:
\be 
\tau > 103\textrm{yr} \times \frac{\hat{f}_\max}{1000\textrm{Hz}}
\left(\frac{\Tcoh}{1800\textrm{s}}\right)^2 \,.
\ee
This restriction will not be present if we use demodulated
data as input for the Hough transform. \\

\noindent \textbf{\emph{Partial and total Hough maps:}}  As described above, 
for a given frequency bin selected at a given time-stamp and for a 
given value of the instantaneous frequency $\hat{f}$, we can find
the set of sky locations which are consistent with the master equation
(\ref{eq:master1}). In other words, every pixel in the sky-patch
either gets selected or rejected and this gives a histogram in the
$(\alpha,\delta)$ plane consisting of ones or zeros; $\alpha$ and
$\delta$ are coordinates on the sky-patch.  Such a
collection of ones and zeros on the sky-patch is called a 
\emph{Partial Hough Map} (PHM).  The number of PHMs required at any
given time depends on the frequency band $\Delta f_b$ that one is
searching over and is given by $\Delta f_b/\delta f
= \Tobs\Delta f_b$.  

Given a set of PHM's for every time interval, and given a set of
spin-down parameters that one wishes to search for, the \emph{Total
Hough Map} (THM) is obtained by summing the appropriate partial Hough maps.  
\begin{figure}
  \begin{center}
  \includegraphics[height=5cm]{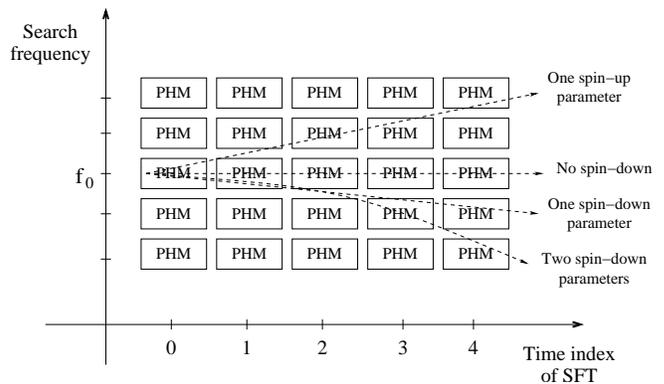}
  \caption{A Partial Hough Map (PHM) is a histogram in the 
  $(\alpha,\delta)$ plane constructed from all the frequencies
  selected at a \emph{given} time and for a \emph{given} value 
  of the instantaneous frequency $\hat{f}_0$.  A total Hough map is
  obtained by summing over the appropriate Hough maps.  The PHMs to be
  summed over are determined by the choice of spin-down parameters
  which give a trajectory in the time-frequency plane.  For
  example, a single spin down parameter will give a straight line as
  shown is the figure while two spin-down parameters will lead to a
  parabola.}\label{fig:sumphm}
  \end{center}
\end{figure}
To see how this comes about, consider the case when we are searching for some
spin-down parameters $\{f_{(n)}\}$ with $n=1,2,\cdots$. The
instantaneous frequency changes with time according to equation
(\ref{eq:f0hat}); ignore the $\Delta \mathbf{r}$ term in this
equation \footnote{This is just
saying that the spin-down age $\tau:=\hat{f}/f_{(1)}$ of the   
star is much larger than the light travel time $|\Delta\mathbf{r}|/c$.   
}. 
This can be viewed as a trajectory in the time-frequency plane.  A
single spin-down parameter will give a straight line, two spin-down
parameters a parabola and so on.  Thus for each time-stamp $t_a$, we
can find the appropriate PHM by looking at which frequency bin this
trajectory intersects (see figure \ref{fig:sumphm}).  For a given
choice of spin-down parameters, the THM 
is obtained by summing over the appropriate
PHMs.  Repeating this for every set of frequency and 
spin-down parameters we wish to
search over, we obtain a number of THMs and the collection of all
these THMs represent our final histogram in parameter
space.\\ 

\noindent \textbf{\emph{Look up tables:}} The procedure described thus far is, in
principle, enough to produce a complete Hough map in parameter space.
However, it is possible to enormously reduce 
the computational cost by
using \emph{Look Up Tables} (LUTs) which we now describe.  Assume that
we have managed to find all the annuli for a given time-stamp $t_a$
and for a given search frequency $\hat{f}$.  To construct the PHM
for $t_a$ and $\hat{f}$, we just need to select the appropriate
annuli out of all the ones that we have found.  Very importantly, it
turns out that in most cases, the annuli are relatively insensitive to
changes in $\hat{f}$ and can therefore be re-used a large number of
times.

To see this, look at how the solutions of
equation (\ref{eq:master1}) depend on the search frequency
$\hat{f}$.  We want to calculate the maximum number $\kappa$ of
frequency bins that $\hat{f}$ can be changed by so that the annuli
change by only a fraction $r$ of the quantity
$(\delta\phi)_{\textrm{min}}$ defined in equation
(\ref{eq:minannulus}).  As discussed earlier, if we restrict ourselves to
discrete frequencies, the annuli corresponding to every given value of
$\hat{f}$ are parameterized by an integer $n$ according to equation
(\ref{eq:annulilabels}). For a fixed value of $n$, by how much does
$\phi_n$ change when $\hat{f}$ is varied?  To answer this,
differentiate equation (\ref{eq:annulilabels}) with respect to $\hat{f}$:
\be
\frac{d\hat{f}}{\hat{f}^2} =
\frac{1}{n\delta f}\frac{v}{c}\sin\phi_n\, d\phi_n =
\frac{\tan\phi_n}{\hat{f}} d\phi_n \,.
\ee
Set $d\phi_n=r(\delta\phi)_\min$ and $d\hat{f} =
\kappa\delta f=\kappa/\Tcoh$ to obtain
\be \label{eq:kappaphin} \kappa = \frac{rc}{v}\tan\phi_n  =
\frac{rc}{v}\sqrt{\frac{n_0^2}{n^2} -1} \ee
where $n_0 = v\hat{f}/(c\delta f)$.  
Consider separately the two regimes when $\phi_n\sim\pi/2$
(i.e. $n\sim 0$) and
$\phi_n\sim 0,\pi$ (i.e. $n\sim \pm n_0$).  When $\phi_n=\pi/2$, then
$\kappa$ is infinite which 
indicates that a LUT is excellent in this regime.  On the other hand,
$\kappa=0$ for $\phi_n=0,\pi$.  However, since the resolution in
$\phi$ is finite, instead of $\phi_n=0$, it is more appropriate to
take the worst case scenario as $\phi_n = (\delta\phi)_{\min}$ so that 
\be \kappa \approx  \frac{rc}{v}\left(\delta\phi\right)_{\textrm{\mbox{\tiny{min}}}} =
40r\left(\frac{500\,\textrm{Hz}}{\hat{f}}\right)^{\frac{1}{2}}\,. \ee
Thus, in this worst case scenario, for a frequency of 500 Hz and a
tolerance of $r=0.1$, the LUT will be valid for $4$ frequency
bins.  Furthermore, due to the presence of the function $\tan\phi$ in
equation (\ref{eq:kappaphin}), $\kappa$ increases rapidly with
increasing $\phi_n$ (i.e. decreasing $n$).  As an example, take
$\Tcoh=1800$s, $\hat{f} = 500$ Hz, and $v/c = 10^{-4}$ so that
$n_0=90$.   Then, even for $n=89$, we get $\kappa=1500r$; thus with
say $r=0.1$, the LUT is valid for about $150$ frequency bins.  

The main point of using the look up tables and partial Hough maps
 is to reduce the computational costs.  
Assume that we are searching over a
frequency band $\Delta f_b$, let $N_p$ be the number of templates
in the space of sky-locations and spin-down parameters;
$\Tcoh \Delta f_b$ is the number of frequency bins being searched over.  

A naive implementation of the Hough transform will require,  
for every point in parameter space, to identify first the corresponding
pattern  in the time-frequency plane and then 
add $N$ integers (zeros or ones) to obtain the final number count,
where $N$ is the number of data segments.
If we use LUTs, which are valid  for a large number of search frequencies,
their computational cost in the search becomes negligible, 
i.e. the cost of finding the patterns is negligible,
 and the number of floating point operations required is thus  $C_0\approx 
\Tcoh\Delta f_bN_pN$.
This calculation can be organized much more efficiently if we perform
a search on many sky locations at once. In this case,
if we know the locations of all the annuli,
for every chosen frequency bin, we mark the corresponding annulus and
in the end, combine all the annuli thus selected to get the final
number count. 
The exact savings in computational cost due to this strategy are implementation
dependent, but are typically better by a factor of $\sim 5$ when compared
to the  value $C_0$ mentioned above.  This factor is related to the number
of frequency bins selected from each SFT.

\section{Statistical properties of the Hough maps}
\label{sec:statistics}

This section is divided into three parts: The probability distribution
of the number counts is calculated in subsection \ref{subsec:statistics},
subsection \ref{subsec:optimalthr} optimizes the various thresholds
and subsection \ref{subsec:sensitivity} estimates the sensitivity of
the Hough search.   

\subsection{The number count distribution}
\label{subsec:statistics}

The frequency bins that are fed into the Hough transform are the ones
such that their normalized power $\rho_k$ defined in equation (\ref{eq:rhodef})
exceeds a threshold $\rho_\th$. 
Assuming that the noise is stationary, has zero mean, and
is Gaussian,
from equation (\ref{eq:rhodef}), we get
\be
2\rho_k = z_1^2+z_2^2
\ee
where
\be
z_1 = \frac{\sqrt{2}\textrm{Re}[\tilde{x}_k]}{\sqrt{\langle
|\tilde{n}_k|^2\rangle}}\qquad\textrm{and} \qquad z_2 =
\frac{\sqrt{2}\textrm{Im}[\tilde{x}_k]}{\sqrt{\langle |\tilde{n}_k|^2
    \rangle}}\,  .
\ee
As before, the detector output $\tilde{x}_k$ is the sum of noise and a
possible signal: $\tilde{x}_k=\tilde{n}_k+\tilde{h}_k$. 
Assuming that $\textrm{Re}[\tilde{n}_k]$ and
$\textrm{Im}[\tilde{n}_k]$ are independent random variables with equal  
variance, it is easy to show that their variance must be equal to
$\langle |\tilde{n}_k|^2\rangle/2$.  Therefore, taking the noise to be
Gaussian, it follows that the random variables $z_1$ and $z_2$ 
are normally distributed and have unit variance (but non-zero mean).
Thus $2\rho_k$ must be distributed according to a non-central $\chi^2$
distribution with 2 degrees of freedom with non-centrality parameter
$\lambda_k$:
\be \label{eq:noncentral}
\lambda_k = \left(\mathbf{E}[z_1]\right)^2 + \left(\mathbf{E}[z_2]\right)^2 =
\frac{4|\tilde{h}(f_k)|^2}{\Tcoh S_n(f_k)} \,.
\ee
Thus the distribution of $\rho_k$ is 
\ba  
p\left(\rho_k|\lambda_k\right) &=&
2\chi^2(2\rho_k|\,2,\lambda_k) \nonumber \\ 
&=& \exp\left(-\rho_k - \frac{\lambda_k}{2}\right)
I_0(\sqrt{2\lambda_k\rho_k})\label{eq:chisquare}
\ea
where $I_0$ is the modified Bessel's function of
zeroth order. As expected, $p\left(\rho_k|\lambda_k\right)$ reduces to 
the exponential distribution in the absence of a signal (when $\lambda=0$). 

\begin{figure}
    \includegraphics[height=5cm]{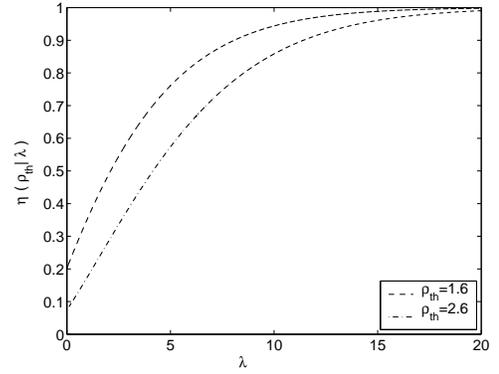}
    \caption{\small{Plot the detection probability
    $\eta(\rho_\th|\lambda)$ as a function of $\lambda$ for
    $\rho_\th=1.6$ and $2.6$. }}\label{fig:detectprob}
\end{figure}

The mean and variance for this distribution are respectively
\be \mathbf{E}[\rho_k] = 1 + \frac{\lambda_k}{2} \qquad\textrm{and}\qquad 
    \mathbf{\sigma^2}[\rho_k] = 1+\lambda_k \, .
\ee
The probability $\eta$ that a given frequency bin is selected is 
\be \label{eq:probeta}
\eta(\rho_{\th}|\lambda) = \int_{\rho_{\th}}^\infty p(\rho|\lambda)\,\textrm{d}\rho
\ee
where we have dropped the subscript $k$ for notational
simplicity; it is understood that $\rho$ and $\lambda$ always refer to
one of the Fourier frequency bins.  
The false alarm and false dismissal probabilities for the frequency
bin selection are respectively
\ba 
\alpha(\rho_{\th}) &=& \int_{\rho_{\th}}^{\infty}
p(\rho|0)\,\textrm{d}\rho =  e^{-\rho_{\th}}\,, \label{eq:powerfa} \\
\beta(\rho_{\th}|\lambda) &=& 1-\eta(\rho_{\th}|\lambda) = \int_0^{\rho_{\th}}
p(\rho|\lambda)\,\textrm{d}\rho \label{eq:powerfd}
\, .
\ea
Clearly, $\eta = \alpha$ when no signal is present and $\eta$ becomes
larger when the signal becomes stronger and $\eta \rightarrow 1$ when 
$\lambda\rightarrow \infty$. Figure \ref{fig:detectprob} shows 
$\eta(\rho_\th |\lambda)$ as a function of the non-centrality parameter
$\lambda$ for two different values of $\rho_\th$.  For small $\lambda$:
\be \label{eq:etasmallsignal}
\eta(\rho_\th|\lambda) = \alpha \left\{1+\frac{\rho_\th}{2}\lambda +
\mathcal{O}(\lambda^2)  \right\} \,.
\ee
This expansion will be very useful when we restrict ourselves to the
case of small signals.

In the presence of a signal, the
non-centrality parameter $\lambda_k$ is \emph{not} constant across
different SFTs. 
The reason for this is two-fold: First, the noise may
be significantly non-stationary.  Secondly, and more
fundamentally, the observed signal power $|\tilde{h}|^2$ changes
because of the amplitude modulation of the signal caused by the
non-uniform antenna 
pattern of the detector. Therefore, the detection probability
$\eta$ changes across SFTs.  In what follows, we shall neglect this
effect and take $\lambda$ and $\eta$ to be constant for different SFTs.  

Under this assumption, the probability of measuring a number count $n$
in a pixel of a Hough map constructed from $N$ SFTs is given by the
binomial distribution:
\be \label{eq:binomial}
p(n|\rho_\th,\lambda) = \left( \begin{array}{c} N \\ n \end{array}  \right)
\eta^n(1-\eta)^{N-n}\, .
\ee
The mean and variance of the number count are respectively
\be
\bar{n} = N\eta \qquad \textrm{and}\qquad \sigma^2 =
N\eta (1-\eta) \, .\label{eq:binomeanvar}
\ee
In the absence of a signal, $\eta=\alpha$ so that
\be \label{eq:binomialnosig}
p(n|\rho_\th,0) = \left( \begin{array}{c} N \\ n \end{array}  \right)
\alpha^n(1-\alpha)^{N-n}\, .
\ee

Candidates for detection or for further analysis
are selected by setting a threshold $n_\th$ on the number count.
Based on this, we can define the false alarm and false
dismissal rates respectively as:
\ba \label{eq:houghalpha}
\alpha_H(n_\th,\rho_\th,N) &=& \sum_{n=n_\th}^{N} p(n|\rho_\th,0) \,,\\
\beta_H(n_\th,\rho_\th,\lambda,N) &=& \sum_{n=0}^{n_\th-1}
p(n|\rho_\th,\lambda)\,.\label{eq:houghbeta} 
\ea
These two quantities determine the significance and the sensitivity of
the Hough search and will play an important role in the rest of
this paper.  

\subsection{Optimal choice of the thresholds}
\label{subsec:optimalthr}

In order to carry out the Hough search, we have to set two
thresholds: the threshold $\rho_\th$ on the normalized power and the
threshold $n_\th$ on the number count.  

The value of $n_\th$ is determined by the false alarm rate
$\alpha_H^\star$ that depends on the number of candidates that we can
feasibly follow up.  

The value of $\rho_\th$ is chosen in such as way so as to make
the search as powerful as possible.  We present two criteria that
yield the same result for small signals and for large $N$. \\

\noindent \textbf{\emph{Maximizing the critical ratio:}} For the Hough
number count, we can define a random variable called the
\emph{critical ratio} as follows
\be
\Psi = \frac{n-N\alpha}{\sqrt{N \alpha(1-\alpha)} } \, ,
\ee
This quantity is a measure of the ``significance'' of a measured value
$n$ with respect to the expected value $N\alpha$ in the absence of any
signal, weighted by the
expected fluctuations of the noise. In the presence of a signal, 
the expected value of the critical ratio is
\be \label{eq:psibar}
\bar{\Psi}(\eta, \alpha) = \frac{N\eta-N\alpha}{\sqrt{N
    \alpha(1-\alpha)} } \, , 
\ee
Recall that $\eta$ and $\alpha$ depend on the threshold $\rho_\th$
according to equations (\ref{eq:powerfa}) and (\ref{eq:powerfd})
respectively.  Thus, our criterion for choosing the threshold
is to maximize $\bar{\Psi}(\eta,\alpha)$ with respect to $\rho_\th$. In the
case of small signals where $\eta\approx\alpha(1+\rho_\th\lambda/2)$,
the condition
\be \label{eq:psimin}
\frac{\partial \bar{\Psi}}{\partial \rho_\th} = 0
\ee
leads to
\be
\ln \alpha = 2(\alpha -1)  
\ee
which yields $\rho_\th \approx 1.6$ or equivalently, $\alpha \approx
0.20$. \\

\noindent \textbf{\emph{The Neyman-Pearson criterion:}}  An
alternative method of choosing $\rho_\th$ is based on the
Neyman-Pearson criterion which tells us to minimize the false
dismissal rate $\beta_H$ for a given value $\alpha_H^\star$ of the
false alarm rate. For weak signals, this requirement uniquely determines
$\rho_\th$ and, as we shall see, this agrees with the previous
criterion.  

In practice, taking $N$ and $\lambda$ to be fixed parameters, this is the procedure:
\begin{description}
\item[i.] First choose a value $\alpha_{H}^{\star}$ for the largest false alarm rate
  $\alpha_H$ that we can allow. 
\item[ii.] Invert the equation
  $\alpha_H(\rho_\th,N,n_\th) \leq \alpha_H^\star$ to 
  obtain $n_\th(\rho_\th,N,\alpha_{H}^{\star})$. 
\item[iii.] Substitute the value of $n_{\th}$ thus obtained in the expression for the
  false dismissal $\beta_H(n_\th,\rho_\th,\lambda,N)$.  This gives $\beta_H$ as a
  function of $(\rho_\th,\lambda,N,\alpha_{H}^{\star})$.  
\item[iv.] Minimize $\beta_H$ as a function of $\rho_\th$. Let
  $\rho_\th^\star$ be the value that minimizes $\beta_H$. 
\item[v.] Using $n_\th(\rho_\th,N,\alpha_{H}^{\star})$ derived in step
  (ii) above, obtain $n_\th^\star=n_\th(\rho_\th^\star,N,\alpha_{H}^{\star})$.
\end{description}
This procedure is also applicable if we choose a different method of
selecting frequency bins other than simple thresholding, such as, for
example the peak selection criterion mentioned towards the end of
subsection \ref{subsec:notation}.  

\begin{figure}
    \includegraphics[height=6cm]{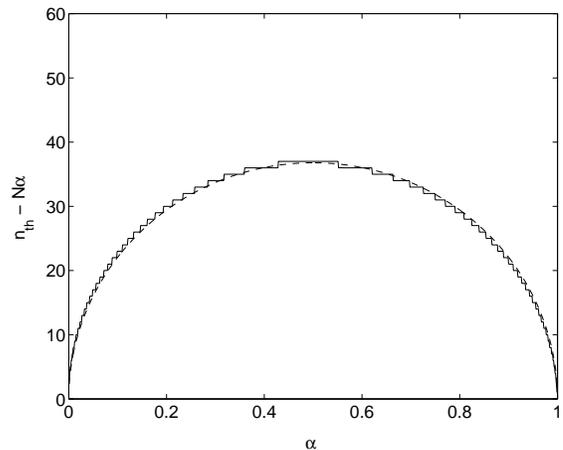}
    \caption{\small{Graph of $n_\th-N\alpha$ versus
    the false alarm probability $\alpha=e^{-\rho_\th}$ for
    $\alpha_H=0.01$ and $N=2000$. The dashed line shows the analytic
    approximation given by equation
    (\ref{eq:nthrapprox}). }}\label{fig:nthr}   
\end{figure}

\begin{figure}
    \includegraphics[height=6cm]{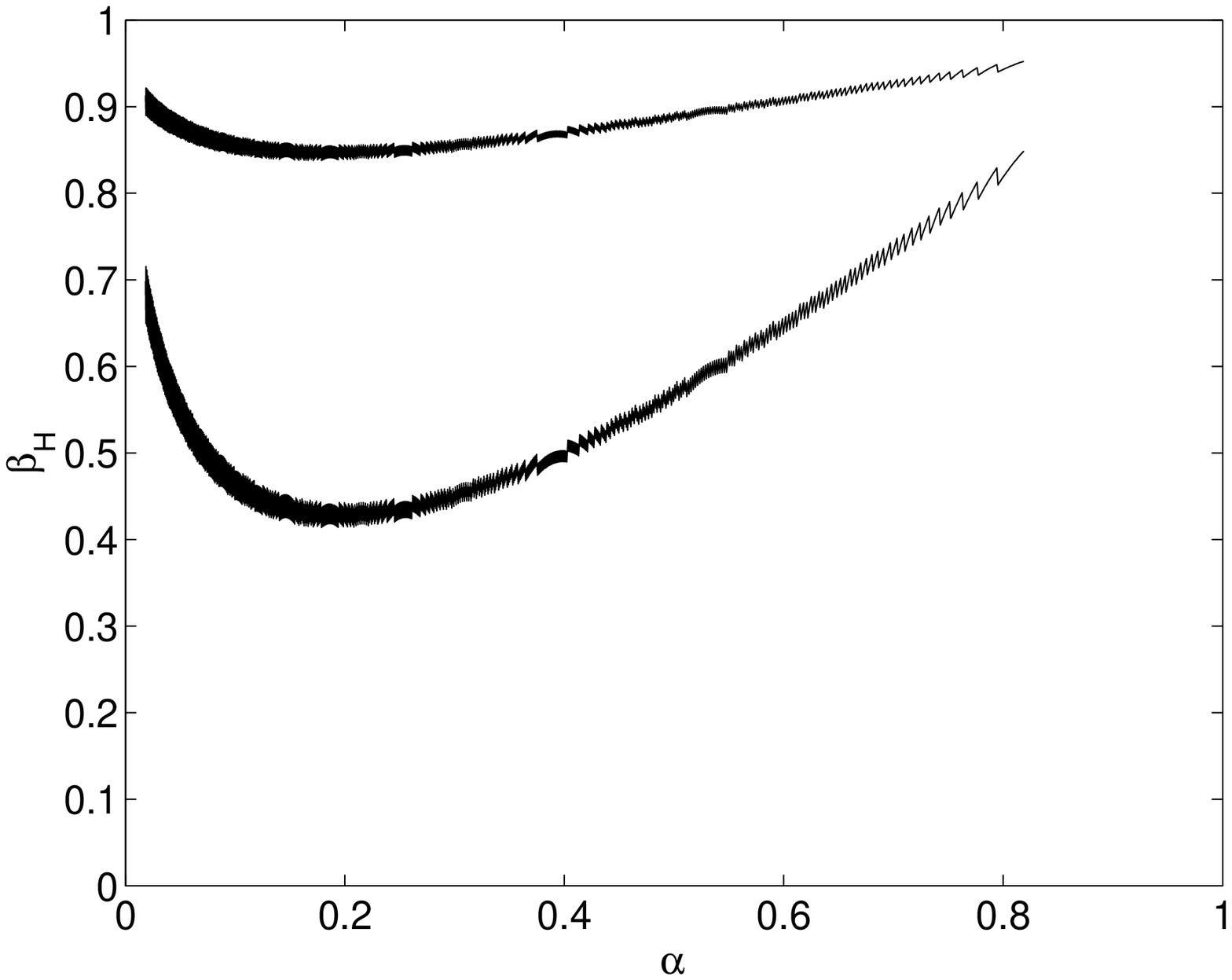}
    \includegraphics[height=6cm]{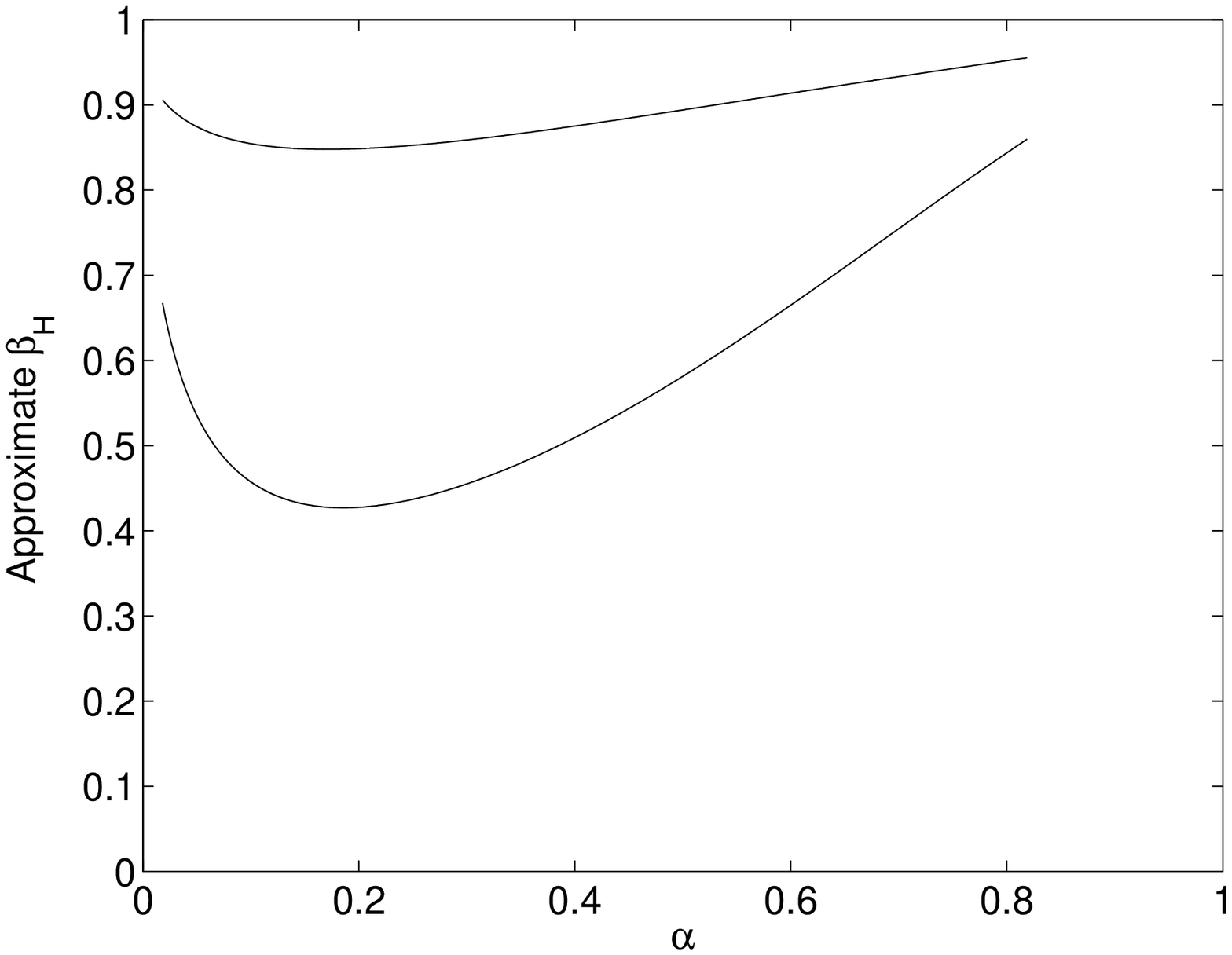}
    \caption{\small{The first figure shows the Hough false dismissal rate as a
    function of the false alarm rate $\alpha=e^{-\rho_\th}$ for a
    non-centrality parameter $\lambda=0.10$ (upper curve) and
    $\lambda=0.20$ (lower curve).  Both curves correspond to
    $\alpha_H=0.01$ and $N=1000$. The minimum values of $\beta_H$ for
    the two curves are approximately $0.84$ and $0.41$ respectively.
    Both minima occur at $\alpha=0.20$ approximately.  This
    corresponds to a threshold of $\rho_\th=1.6$ on the normalized
    power statistic. The bottom figure shows the approximation to
    $\beta_H$ using equations (\ref{eq:betaHapprox}) and
    (\ref{eq:nthrapprox}) with the same parameters as in the first
    figure.  }}\label{fig:minbeta}    
\end{figure}

\begin{figure}
    \includegraphics[height=6cm]{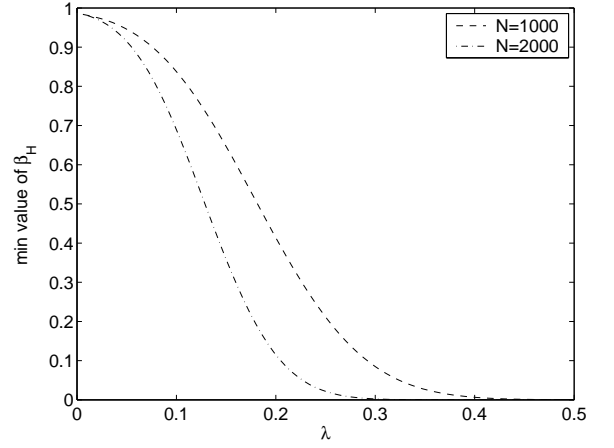}
    \caption{\small{Minimum value of $\beta_H$ as a function of the
    non-centrality parameter $\lambda$ for $\alpha_H=0.01$ and for
    $N=1000$ and $2000$. As expected, a larger value of $N$ typically
    leads to a smaller value of $\beta_H$. }}\label{fig:minvalbeta}  
\end{figure}

The results of the optimization procedure described above are shown
in figures \ref{fig:nthr}, \ref{fig:minbeta} and \ref{fig:minvalbeta}.
Figure \ref{fig:nthr} shows the value of the number count threshold
$n_\th$ obtained as described in step (ii).
In this figure, instead of $\rho_\th$,
we have chosen the false alarm rate $\alpha=e^{-\rho_\th}$ as the
independent variable; $\alpha$ is the false alarm rate for selecting
frequency bins and is not to be confused with $\alpha_H$.  Figure
\ref{fig:nthr} also shows an analytic approximation to $n_\th$
obtained below in equation (\ref{eq:nthrapprox}).  Using this result
for $n_\th$, figure \ref{fig:minbeta} shows $\beta_H$ as a function
of $\alpha=e^{-\rho_\th}$.  The optimal choice $\rho_\th^\star$ of
$\rho_\th$ is when $\beta_H$ is a   
minimum and, for small signals, this happens at 
$\rho_\th^\star\approx 1.6$ which corresponds to
$\alpha^\star:=e^{-\rho_\th^\star}\approx 0.20$ .
Finally, figure \ref{fig:minvalbeta} shows the minimum 
value of $\beta_H$ obtained by this optimization as a function of the
signal strength $\lambda$ and for two different values of $N$.  

\noindent \textbf{\emph{The Gaussian approximation:}} To better understand
the statistics, it is useful to carry out the above 
steps analytically by taking $n$ to be a continuous variable
and by approximating the binomial distribution by a Gaussian with the
appropriate mean and variance: 
\be 
p(n|\rho_\th,\lambda) =
\frac{1}{\sqrt{2\pi\sigma^2}}e^{-(n-N\eta)^2/2\sigma^2}\,.  
\ee
This is a very good approximation when $N$ is large and $\eta$ is not
too close to $0$ or $1$.  If $n$ is chosen to lie within
$[0,N]$, the distribution is properly normalized only
approximately.  For simplicity, in what follows we shall take the
range of $n$ to be $(-\infty,+\infty)$; this is an excellent
approximation if the above assumptions on $N$ and $\eta$ hold.

With the approximations given above, we can rewrite the equation
$\alpha_H=\alpha_{H}^{\star}$ as 
\be
\int_{n_\th}^\infty p(n|\rho_\th,0)\,dn = \alpha_{H}^{\star}\,.
\ee
The solution to this equation can be rewritten in terms of the complementary
error function: 
%
\be \label{eq:nthrapprox}
n_\th(\rho_\th,N,\alpha_{H}^{\star}) = N\alpha +
\sqrt{2N\alpha(1-\alpha)}\,\,\textrm{erfc}^{-1}(2\alpha_{H}^{\star}) \,.
\ee
As shown in figure \ref{fig:nthr}, this is a very good approximation
to the actual value of $n_\th$ obtained from the binomial
distribution.  

The expression for $\beta_H$ is similarly rewritten as
\be \label{eq:betaHapprox}
\beta_H =  \frac{1}{2}\textrm{erfc}\left(
\frac{N\eta-n_\th}{\sqrt{2N\eta(1-\eta)}}\right) \,.\nonumber
\ee
As figure \ref{fig:minbeta} shows, this too is a very good approximation
to $\beta_H$ obtained using the binomial distribution.

In step (iv), we find $\rho_\th^\star$ such that 
\be 
\left.\frac{\partial \beta_H}{\partial \rho_\th}\right|_{\rho_\th=\rho_\th^\star} = 0
\ee
In the case of small signals where
$\eta\approx\alpha(1+\rho_\th\lambda/2)$, the solution to the above
equation becomes independent of $\lambda$, and it is also
independent of $N$ when $N$ is large.  In these limits, the solution
is given by
\be
\left.\frac{\partial}{\partial \rho_\th}\left(\sqrt{\frac{\rho_\th^2}{e^{\rho_\th}-1}} \right)\right|_{\rho_\th=\rho_\th^\star}=0\,.
\label{D4.53}
\ee
The solution to this equation is $\rho_\th^\star\approx 1.6$ and
$\alpha^\star=e^{-\rho_\th^\star}\approx 0.20$.  Notice that this equation
is equivalent to equation (\ref{eq:psimin}) and furthermore, the
functions being extremized are rather flat near the extremum.  Thus, the
threshold could be chosen somewhat differently without significantly
impacting the sensitivity.  In particular, the threshold can be  
increased so that fewer frequency bins are selected.  Depending
on the details of the implementation, this could lead to a lower
computational cost; in the framework of a hierarchical search, this
will improve the overall sensitivity.  

Finally, with the optimal threshold $\rho_\th^\star$ at hand, the
optimal threshold $n_\th^\star$ on the number count is obtained by
substituting $\rho_\th=\rho_\th^\star$ in equation
(\ref{eq:nthrapprox}):
\ba n_\th^\star &=& n_\th(\rho_\th^\star,N,\alpha_H^\star) \\
&=& N\alpha^\star +
\sqrt{2N\alpha^\star(1-\alpha^\star)}\,\textrm{erfc}^{-1}(2\alpha_H^\star)\,.
\nonumber  \ea
This is an important equation because
it tells us the number count threshold that must be set in order
to have a given number of follow-up candidates.

\subsection{Sensitivity}
\label{subsec:sensitivity}

In this subsection, we estimate the sensitivity of the Hough search,
i.e. we answer the following question: for given values $\alpha_{H}^{\star}$ and
$\beta_H^{\star}$ of the false
alarm $\alpha_H$ and false dismissal $\beta_H$ respectively, what is the
smallest value of the gravitational wave amplitude $h_0$ (see equation
(\ref{eq:h0})) that would cross the thresholds $\rho_\th$ and $n_\th$?
Equivalently, for a given false alarm rate $\alpha_{H}^{\star}$, what is the
smallest $h_0$ which will give a false dismissal rate of at least $\beta_H^{\star}$?
We use the signal model of equation (\ref{eq:amplitude2}) and we
present our final result for the values
$\alpha_H=\alpha_{H}^{\star}=0.01$ and  
$\beta_H=\beta_H^{\star}=0.10$.  The value of $0.01$ is meant mainly
for illustration purposes and does not change the results
qualitatively.  Furthermore, for comparison, equation (2.2) in
\cite{S1:pulsar} assumes a false alarm of $0.01$ and this choice of
$\alpha_H^{\star}$ enables an easier comparison with that result.  As far as
possible, we explicitly retain the factors of $\alpha_H^{\star}$ in
our equations substituting numerical values only when necessary.    

We must first solve the equation 
\be \beta_H(n_\th^\star, \rho_\th^\star, \lambda, N) = \beta_H^{\star}  \ee
and obtain $\lambda$ as a function of $N$; this will yield the
desired value of $h_0$. 

In order to simplify the discussion, we shall again approximate the
binomial distribution by a normal
distribution whose mean $\bar{n}$, and variance $\sigma$, are
respectively given by equation (\ref{eq:binomeanvar}).
The false dismissal rate is 
\be\label{eq:solvebeta}
\beta_H \approx \frac{1}{2}\textrm{erfc}\left(
\frac{\bar{n} - n_\th^\star}{\sqrt{2N\eta^\star(1-\eta^\star)}}\right) 
\ee
where $n_\th^\star$ is as given in equation (\ref{eq:nthrapprox}) and
$\eta^\star=\eta(\rho_\th^\star|\lambda)$. 

\begin{figure}
    \includegraphics[height=5cm]{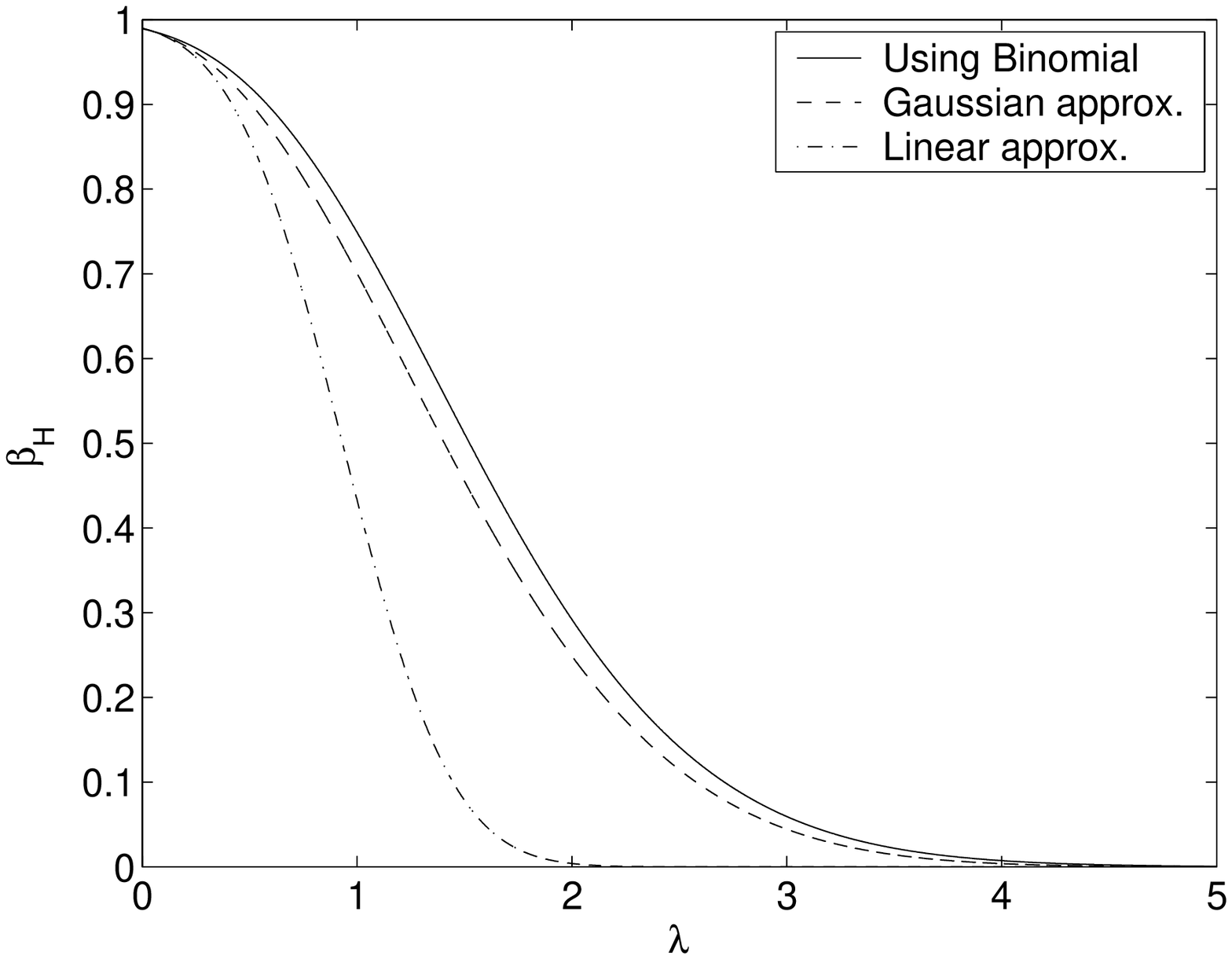}
    \includegraphics[height=5cm]{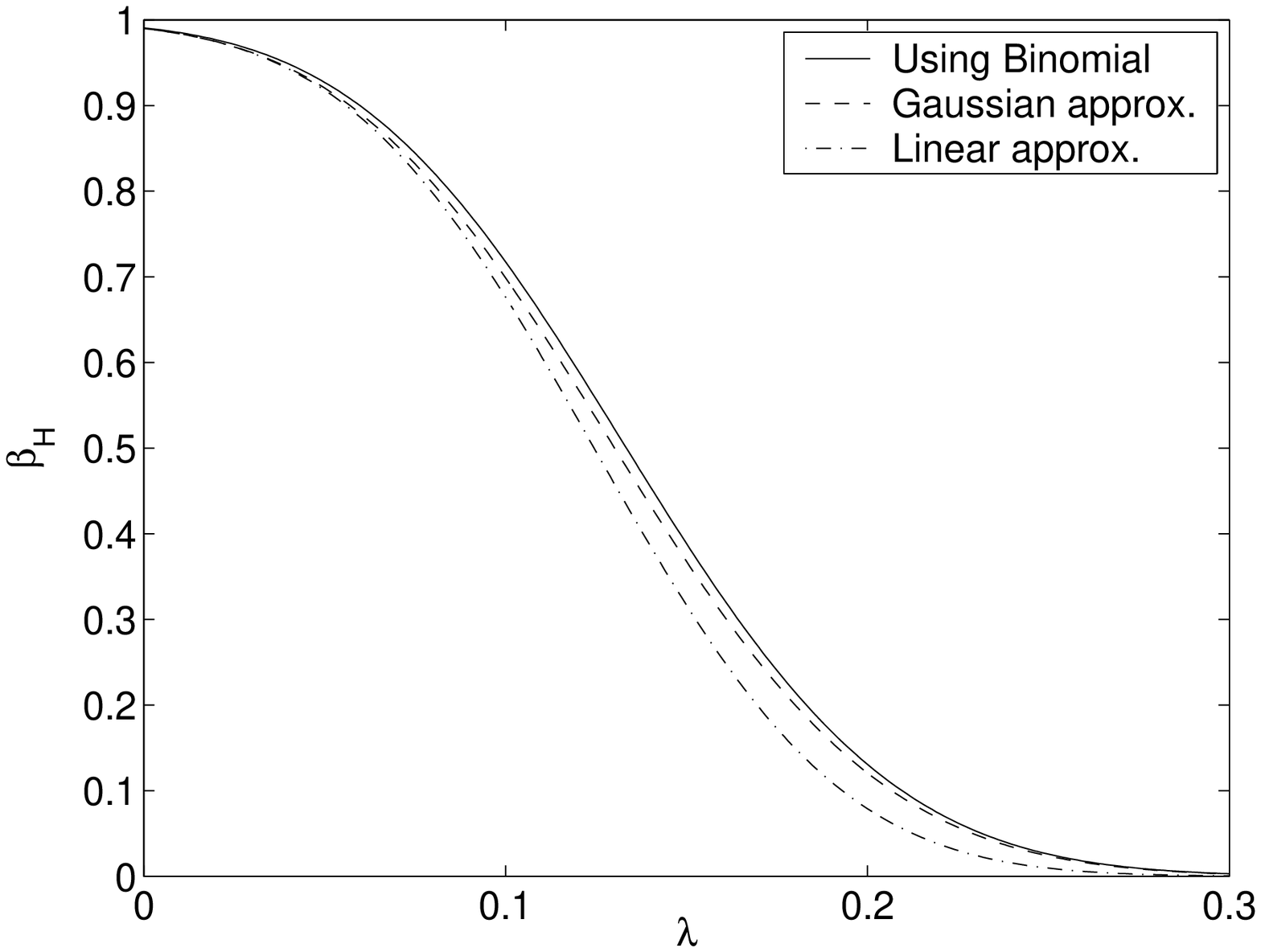}
    \includegraphics[height=5cm]{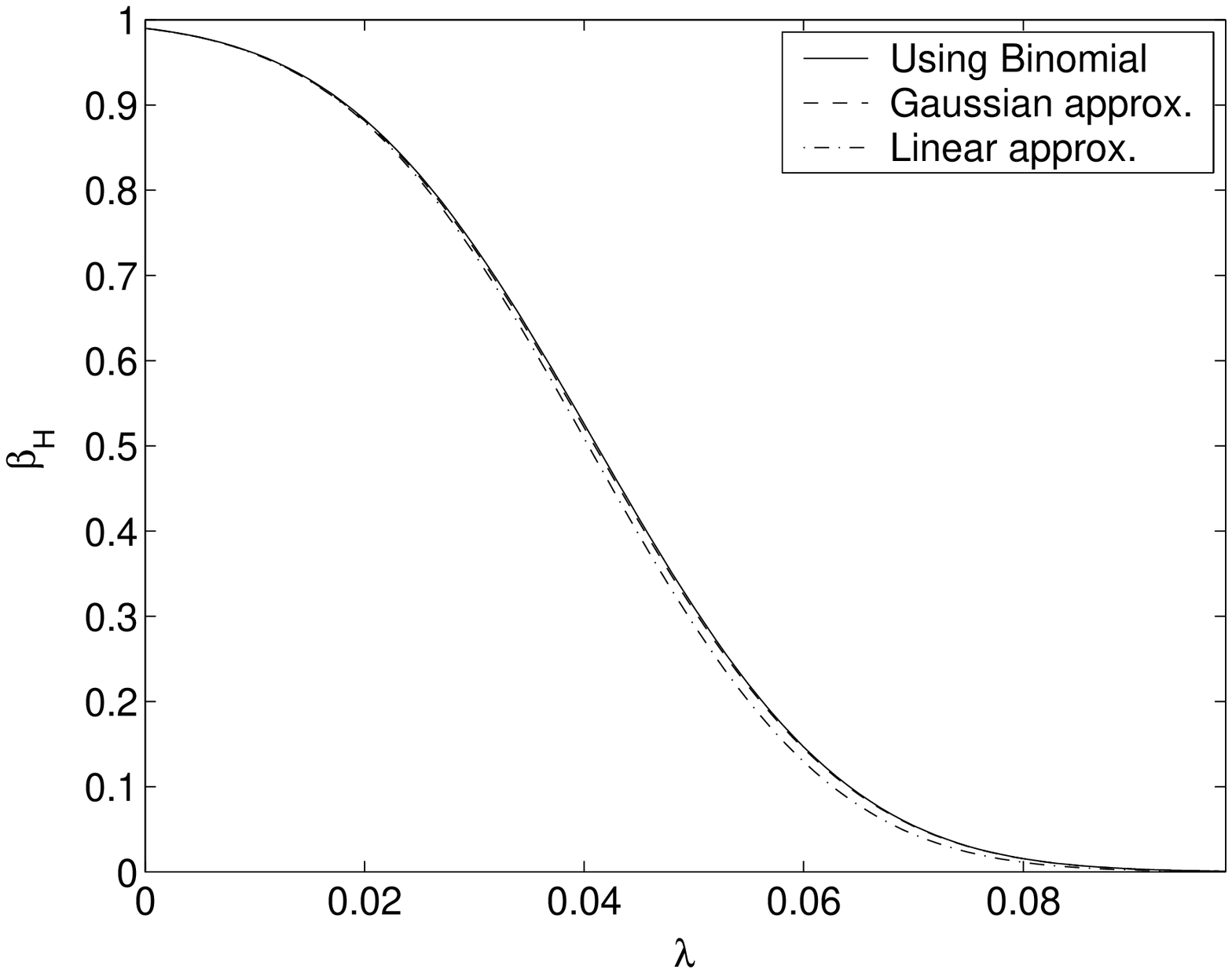}
    \caption{\small{Graphs of $\beta_H$ as a function of $\lambda$ for
    different values of $N$ and for the three different approximations
    used. In the first panel $N=20$, the second panel has $N=2000$ and
    in the third panel, $N=20,000$.  All graphs are plotted assuming
    the optimal values for $\rho_\th$ and $n_\th$. The linear approximation
    is clearly unacceptable for $N\sim 10^1$ but becomes reasonable when $N\sim
    10^2$ or $10^3$ and is excellent for $N\sim
    10^4$. The Gaussian approximation is clearly much better and is good even
    for $N=20$.  Finally, note that the approximations always underestimate
    the value of $\beta_H$. }}\label{fig:minvalbetaN}   
\end{figure}

Since we are interested in the case of small signals, let us
approximate $\eta$ by only keeping terms of the order of $\lambda$ in
equation (\ref{eq:etasmallsignal}).  Ignoring terms of
$\mathcal{O}(\lambda^2)$, equation (\ref{eq:solvebeta})
leads to the approximation 
\be\label{eq:solvebetaapprox}
\beta_H = \frac{1}{2}\textrm{erfc}\left(-\textrm{erfc}^{-1}(2\alpha_{H}^{\star})
+ \frac{1}{2} \Theta \alpha^\star\rho_\th^\star\lambda\right)
\ee
where
\be \label{eq:c}
\Theta = \sqrt{\frac{N}{2\alpha^\star(1-\alpha^\star)}} +
\left(\frac{1-2\alpha^\star}{1-\alpha^\star}\right)\frac{\textrm{erfc}^{-1}
  (2\alpha_{H}^{\star})}{2\alpha^\star} \,.
\ee 
Let us summarize our approximation scheme for $\beta_H$. 
The first approximation
is to take the number count distribution to be binomial.
The second approximation is in equation (\ref{eq:solvebeta}) which
replaces the binomial by a Gaussian distribution with the appropriate
mean and variance.  The final approximation is in equation
(\ref{eq:solvebetaapprox}) where we have taken $\lambda$ to be
small and used a Taylor series in powers of $\lambda$ retaining only
the linear term.  To get a feeling for the validity of these
approximations, figure \ref{fig:minvalbetaN} shows graphs of $\beta_H$
as a function of $\lambda$ for different values of $N$.  As the graphs
show, we can trust the approximations when $N\sim10^3$.  For smaller
values of $N$, while the Gaussian approximation is still reasonable, 
the linear approximation greatly under-estimates $\beta_H$ for a given
value of $\lambda$, i.e. it make the Hough search appear more
sensitive than it actually is.  

Working with the linear approximation of equation
(\ref{eq:solvebetaapprox}), assuming $N$ to be very large and
$\textrm{erfc}^{-1}(2\alpha_{H}^{\star}) \ll N$, set 
$\beta_H=\beta_H^{\star}$ and solve for $\lambda$:  
\be \label{eq:lambdaapprox}
\lambda \approx
\frac{\mathcal{S}}{\rho_\th^\star}\sqrt{\frac{8(1-
    \alpha^\star)}{N\alpha^\star}} 
\approx
\frac{9.02}{\sqrt{N}} 
\ee
where 
\be\label{eq:S}
\mathcal{S}:=\textrm{erfc}^{-1}(2\alpha_{H}^{\star}) +
\textrm{erfc}^{-1}(2\beta_H^{\star})
\ee  
and to obtain numerical values, we have chosen
$\alpha_{H}^{\star}=0.01$ and $\beta_H^{\star}=0.10$. Using the
properties of the complementary error function, it is easy to show that
$\mathcal{S}=0$ implies that the statistical significance
$s:=1-\alpha_{H}^{\star}-\beta_H^{\star}$ also vanishes. Therefore,
the quantity $\mathcal{S}$ can be taken to be a measure of the
statistical significance of the search.  The value of $\lambda$
obtained in equation (\ref{eq:lambdaapprox}) gives us the strength of
the smallest signal that can be detected by the Hough search with a
false alarm rate of $1\%$ and a false dismissal rate of $10\%$.   

\begin{figure}
    \includegraphics[height=6cm]{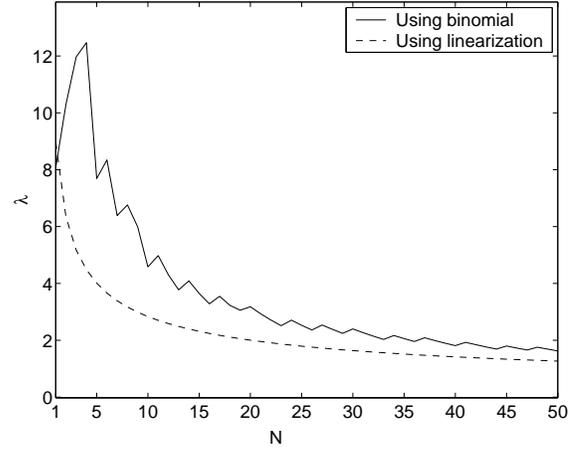}
    \caption{\small{Graph of the smallest detectable $\lambda$ with
    the optimal thresholds. The dashed curve uses the linear
    approximation of equation (\ref{eq:solvebetaapprox}) while the
    solid curve uses the binomial distribution. See text for
    additional discussion.  }}\label{fig:lambda_N}   
\end{figure}

A graph of $\lambda$ as a function of $N$ for small values of $N$ is
shown in figure (\ref{fig:lambda_N}); this figure shows the results
using both the linear approximation and the more accurate binomial
distribution.  The small $N$ limit requires a brief explanation.  For
small $N$, the discrete nature of $n$ becomes important.  In
particular, the false alarm $\alpha_H$ defined in equation
(\ref{eq:houghalpha}) can take only a discrete number of values, the
smallest of which is $\alpha^N$ (at  $n_\th=N$).  Thus for $N=1$, it
is not possible to reach the desired $1\%$ false alarm rate and the
best we can do, with $\rho_\th = 1.6$, is $\alpha_H=0.2$.  To find the
value of $\lambda$ which yields $\beta_H=0.1$, note that for $N=1$,
$\beta_H=1-\eta$. Thus $\eta=1-0.1=0.9$ which implies $\lambda \approx
8.08$; this is the sensitivity of the search for $N=1$.  It
corresponds to a false alarm rate of $20\%$ and a false dismissal rate
of $10\%$. Similar calculations show that the sensitivity becomes worse
as $N$ is increased from $1$ to $4$ as the corresponding false alarm
rates become better. It is only at $N=5$ that we can choose $n_\th<N$
from which point onwards the sensitivity begins to improve.  This  
explains the small $N$ behavior of figure
\ref{fig:lambda_N}.  Similarly, the other discrete jumps in  figure 
\ref{fig:lambda_N} are due to the discrete nature of $\alpha_H$ and
requirement of keeping it below the $1\%$ level.        

To recast the expression for $\lambda$ directly in terms of the signal
amplitude, start with equation (\ref{eq:noncentral}); $\lambda$
depends on the various pulsar parameters.  The
relevant quantity for the purposes of this subsection is the
\emph{average} of $\lambda$ over these 
parameters.  It is quite straightforward to estimate this
average. First, recall the expression for $h(t)$:
\be
h(t) = F_+(t)A_+\cos\Phi(t) + F_\times(t) A_\times\sin\Phi(t) \,.
\ee
Since $\Tcoh$ is much lesser than a day (see equation (\ref{eq:tcoh}))
we can take $F_{+\times}$ to be roughly constant.  Similarly, assuming
that $\Tcoh$ is small enough so that the pulsar signal does not shift
by more than half a frequency bin, we can take the signal frequency
$f(t)$ to be roughly constant. With these approximations, we get
\be
\tilde{h}(f_k) \approx \frac{\Tcoh}{2}(F_+A_+ + F_\times
A_\times)\frac{\sin[\pi (f-f_k)\Tcoh]}{\pi (f-f_k)\Tcoh}
\ee
where $f$ is the instantaneous frequency of the signal and $f_k$ is
the central Fourier frequency of the frequency bin containing $f$; $f$ is
allowed to lie in the range $(f_k-\delta f/2,f_k+\delta f/2)$.  Now
take the square of $\tilde{h}_k$ and average over time to get the
average non-centrality parameter for all the SFTs and note that the
time averages $F_+^2$ and $F_\times^2$ are both $1/5$, and the time
average of $F_+F_\times$ vanishes.  Thus:
\be
\lambda \approx \frac{\Tcoh}{10S_n}(A_+^2 + A_\times^2)
\left(\frac{\sin[\pi (f-f_k)\Tcoh]}{\pi (f-f_k)\Tcoh}\right)^2 \,.
\ee
Take the amplitudes to be of the form given in equations
(\ref{eq:amplitude2}) and average over $\cos\iota\in(-1,1)$ and over the
values of the signal frequency $f\in(f_k-\delta f/2,f_k+\delta f/2)$:
\ba
\langle \lambda\rangle_{\iota,\psi,f,\alpha,\delta} &\approx&
\frac{4}{25}\frac{h_0^2\Tcoh}{S_n}\left(
\int_{-\frac{1}{2}}^{\frac{1}{2}}\frac{\sin^2(\pi x)}{(\pi
x)^2}\,dx\right) \nonumber \\ &\approx& 0.7737\times\frac{4}{25}\frac{h_0^2\Tcoh}{S_n}\,.
\ea
We get the following value for the smallest signal that can
be detected by the Hough search:
\be \label{eq:sensitivity}
h_0=\frac{8.54}{N^{1/4}}\sqrt{\frac{S_n}{\Tcoh}}=8.54N^{1/4}\sqrt{\frac{S_n}{\Tobs}}\,.
\ee
Here the second equality assumes $\Tobs =N\Tcoh$ which be true only if
the $N$ different data segments were contiguous; this is done only to
compare this result with equation (\ref{eq:cohsensitivity}) below. 

Equation (\ref{eq:sensitivity}) is the result we were looking for.
This tells us that if we wish 
to detect a signal with a false alarm rate of $1\%$ and a false
dismissal rate of $10\%$, the weakest signal that will cross the
optimal thresholds is the $h_0$ given above.  The important feature to
note is that $h_0$ is proportional to 
$N^{1/4}/\sqrt{\Tobs}$ while for a coherent search over the whole 
observation time, the sensitivity is proportional to
$1/\sqrt{\Tobs}$.  In particular, for the same values of the false
alarm and false dismissal rates as above, the sensitivity of a full
coherent search directed at around a single point in parameter space
is given by (see~\cite{S1:pulsar}):
\be \label{eq:cohsensitivity}
h_0 = 11.4 \sqrt{\frac{S_n}{\Tobs}} \,.
\ee 
This illustrates the loss in sensitivity introduced 
by combining the different SFTs incoherently but, of course, this is 
compensated by the lesser computational requirements for the
incoherent method.  Furthermore, for say $N=2000$, the sensitivity of
the Hough search is only about a factor of $4.5$ worse than a full
directed coherent search.

This result helps one to make trade-offs of coherent against hierarchical 
searches. For example, if one is searching for a population of objects
that is uniformly distributed in a plane, such as a population of young 
pulsars in the Galaxy, then a coherent search of any region of 
parameter space would go 4.5 times deeper than the incoherent method 
with $N=2000$. The volume of space surveyed in the plane would be
$4.5^2=20$ times  
larger. However, if the incoherent method's speed of execution allowed
it to survey more than $20$ times as much parameter space (including sky 
area and spin-down range) then one would choose the incoherent method. 
This is indeed the case for pulsar searches. 

Finally, equation (\ref{eq:sensitivity}) also allows us to estimate
the astrophysical range of the search.  Combining (\ref{eq:h0}) and
(\ref{eq:sensitivity}), we get:
\ba \label{eq:range}
d &=& \frac{16\pi^2GN^{1/4}I_{zz}\epsilon
f_r^2}{8.54c^4}\sqrt{\frac{\Tcoh}{S_n(f_r)}} \\  
&=& 5.8\,\textrm{kpc}\times \left(
\frac{N}{17000}\right)^{\frac{1}{4}} \left(
\frac{I_{zz}}{10^{38}\textrm{kg-m$^2$}}\right)
\left(\frac{f_r}{500\textrm{Hz}} \right)^2 \nonumber \\
&&\times\left( \frac{\epsilon}{10^{-6}}\right)
\left(\frac{\Tcoh}{1800\textrm{s}}\right)^{\frac{1}{2}}\left(
\frac{10^{-46}\textrm{Hz}^{-1}}{S_n}\right)^{\frac{1}{2}} \nonumber \,.
\ea
Here the reference values for $\Tcoh$ and $N$ have been chosen such
that $N\Tcoh \approx 1$yr, and we have
taken the detector sensitivity to be $10^{-23}\textrm{Hz}^{-1/2}$ at a
frequency of $500$Hz, which is appropriate for the proposed advanced
LIGO detector~\cite{ct}.  A full directed coherent search does
better than this by a factor of about $\sim (17000)^{1/4} \approx
12$. For the initial LIGO detector, assuming it is 10 times less
sensitive than advanced LIGO, we see that it does worse than this by
about a factor of $\sim 3$.

\section{Hough transform with demodulated data}
\label{sec:demod}

As equation (\ref{eq:tcoh}) shows, using the Hough transform with
SFTs as input, necessarily limits the coherent time-baseline $\Tcoh$,
and therefore also the sensitivity of the search.  To get around this
limitation, we need to demodulate each coherent data segment to
remove the frequency drifts caused by the Doppler modulation and the
spin-down;  the only limitation on $\Tcoh$ is then due to the available
computational resources.  The demodulation procedure we use is based
on the $\F$-statistic introduced in~\cite{jks}.  The search pipeline
is very similar to the pipeline shown in figure \ref{fig:nondemod},
the only difference being that instead of computing the power
spectrum, we calculate the $\F$-statistic. Subsection 
\ref{subsec:fstatistic} provides a brief description of the
$\F$-statistic, the master equation is derived in subsection
\ref{subsec:masterdemod}, subsection \ref{subsec:implementdemod} provides
the implementation details and subsection \ref{subsec:demodstatistics}
describes the statistics.

\subsection{The $\F$-statistic}
\label{subsec:fstatistic}

Let $x(t)$ be the calibrated detector output and let $h(t)$ the
waveform that we are searching for.  
In order to extract the signal $h(t)$ from the noise, the optimal
search statistic is the likelihood function $\Lambda$ defined by 
\be \label{eq:likelihood} \ln \Lambda = (x|h)-\frac{1}{2}(h|h) \ee
where the inner product $(\cdot |\cdot)$ is defined as
\be \label{eq:innerprod}
(x|y) := 2 \int_0^\infty \frac{\tilde{x}(f)\tilde{y}^\star(f) +
  \tilde{x}^\star(f)\tilde{y}(f)}{S_n(f)} \, \textrm{d}f \,.\ee
Here, as before, $\tilde{x}(f)$ is the Fourier transform of $x(t)$ and $S_n(f)$ is
the one-sided power spectral density.  The expected
waveform $h(t)$ is given by equations (\ref{eq:waveform}),
(\ref{eq:pluscross}) and (\ref{eq:phasemodel}).  The quantity
$\ln\Lambda$ is essentially the matched filter and is precisely what
we should use in order to best detect the waveform $h(t)$.  However, apart
from the parameters $\vec{\xi}=(f_{(0)},\{f_{(n)}\},\mathbf{n})$,
$\ln\Lambda$ also depends upon the other 
parameters such as the orientation of the pulsar, the polarization
angle of the wave etc. The $\F$-statistic eliminates these additional
variables and enables us to search over only the shape parameters
$\vec{\xi}$.  

Following the
notation of~\cite{jks}, the dependence of the antenna patterns
$F_{+,\times}$ on the polarization angle $\psi$ are given by
\ba
F_+(t) &=& \sin\zeta [a(t)\cos 2\psi + b(t)\sin 2\psi] \\
F_\times (t) &=& \sin\zeta [b(t)\cos 2\psi - a(t)\sin 2\psi]
\ea
where the functions $a(t)$ and $b(t)$ are independent of $\psi$ and
$\zeta$ is the angle between the arms of the detector. 
If we write the phase of the gravitational wave as 
\be
\Phi(t) = \phi_0 + \phi(t)\,,
\ee
then we can always decompose the total waveform $h(t)$ in terms of
four quadratures as
\be
h(t) =\sum_{i=1}^{4} A_i h_i(t)
\ee
where the four amplitudes $A_i$ are time independent and the $h_i$ are
as follows: 
\ba 
h_1(t) &=& a(t)\cos\phi(t)\,, \qquad h_2(t) = b(t)\cos\phi(t)\,, \nonumber\\
h_3(t) &=& a(t)\sin\phi(t)\,, \qquad h_4(t) = b(t)\sin\phi(t)\,. \label{eq:quadratures}
\ea
What this decomposition achieves is a separation of the shape
parameters $\vec{\xi}$ from the other pulsar parameters.
The \emph{only} unknown
parameters in the quadratures $h_i$ are the shape parameters
$\vec{\xi}$ while the amplitudes $A_i$ are independent of
$\vec{\xi}$.  The log likelihood function depends quadratically on
the four amplitudes and we can analytically find the maximum
likelihood (ML) estimators $\hat{A}_i$ of the amplitudes $A_i$ by
solving the set of four coupled linear equations
\be
\left.\frac{\partial \ln\Lambda}{\partial A_i}\right|_{A_i=\hat{A}_i} = 0\,, \qquad i=1,\ldots,4\,.
\ee
The $\F$-statistic is then defined as the log likelihood ratio with the
values of the amplitudes $A_i$ replaced by their ML estimators:
\be
\F := \left.\ln\Lambda \right|_{A_i=\hat{A}_i} \,.
\ee
The only unknown parameters in the optimal search statistic $\F$ are
the shape parameters $\vec{\xi}$.

\subsection{The master equation}
\label{subsec:masterdemod}

Equation (\ref{eq:master1}) describes the expected time-frequency 
pattern when the search statistic is the Fourier transform; in other
words, if the detector output $x(t)$ contains a true signal with
instantaneous frequency $\hat{f}(t)$, then equation (\ref{eq:master1})
tells us the value of the observed frequency $f(t)$ which would
maximize $|\tilde{x}(f)|^2$ in the absence of noise.  If we now use the
$\F$-statistic instead of the Fourier transform, the expected
time-frequency pattern is, as described below, different.  

Before proceeding further, it is useful to
distinguish the instantaneous frequency $f_{(0)}$ from the other shape
parameters which we denote by $\vec{\lambda}$: $\vec{\xi} =
(f_{(0)},\vec{\lambda}) = (f_{(0)},\{f_{(n)}\},\mathbf{n})$.  Let us
the assume that the $\F$-statistic has been computed using the parameters
$\vec{\lambda}_d$ but that the detector output consists of a
signal with parameters $(f_{(0)},\vec{\lambda}$); let us denote
the mismatch in the parameters by $\Delta\vec{\lambda} :=
\vec{\lambda}-\vec{\lambda}_d$ and $\Delta f := f-f_{(0)}$.  

Since the $\F$-statistic is maximized
when the source parameters are equal to the demodulation
parameters, it is clear that if $\Delta\vec{\lambda}=0$, then the
expected time-frequency pattern is just a constant frequency
$f=f_{(0)}$.   More generally, due to the correlations in parameter
space, the mismatch $\Delta\vec{\lambda}$ may produce a residual shift
in the frequency $\Delta f$.  This
frequency shift is determined by
\be \label{eq:maxF}
\left.\frac{\partial \F(f,\vec{\lambda}_d;
f_{(0)},\vec{\lambda})}{\partial
f}\right|_{\vec{\lambda}_d,f_{(0)},\vec{\lambda}} = 0\,.
\ee
Expand $\F$ in powers
of $\Delta\vec{\lambda}$ and $\Delta f$ around the point
$(f_{(0)},\vec{\lambda})$ up to second order (repeated indices are
summed over):
\ba
\F(f,\vec{\lambda}_d;f_{(0)},\vec{\lambda}) &=&
\F(f_{(0)},\vec{\lambda}) + A_{00}\left(\Delta f\right)^2  \\
&+& A_{0i}\Delta\lambda_i\Delta f +
A_{ij}\Delta\lambda_j\Delta\lambda_j \,.\nonumber
\ea
The linear terms do not appear because $\F$ is maximized when $\Delta
f = 0$ and $\Delta\vec{\lambda} = 0$.  With this approximation,
equation (\ref{eq:maxF}) leads to the master equation
\be
\Delta f = -\frac{A_{0i}}{2A_{00}}\Delta\lambda_i
\ee
In other words, the frequency value that 
maximizes the $\F$-statistic for a given $\Delta\vec{\lambda}$, does
not correspond to the intrinsic source frequency $f_{(0)}$ but is
instead given by a linear combination of the $\Delta\lambda_i$.

Let us rewrite equation (\ref{eq:maxF}) more explicitly.  As shown in
\cite{jks}, $\F$ can be written in terms of the amplitude modulation
functions $a(t)$ and $b(t)$ as
\be
\F = \frac{4}{\Tcoh S_n(f_{(0)})}\frac{B|F_a|^2 + A|F_b|^2 - 2C\mathcal{R}(F_aF_b^\star)}{D}
\ee
where $A$, $B$, $C$, and $D$ are constants and 
\begin{eqnarray}
F_a &=& \int_{-\Tcoh/2}^{\Tcoh/2} x(t)a(t)e^{-i\Phi(t;f,\vec{\lambda}_d)}\,dt \, ,\\
F_b &=& \int_{-\Tcoh/2}^{\Tcoh/2} x(t)b(t)e^{-i\Phi(t;f,\vec{\lambda}_d)}\,dt
\, .
\end{eqnarray}
Since we are interested in calculating the frequency drift and not the
amplitude, the variation in the phase is more important than
the amplitude modulation. Thus, the factors of $a(t)$ and $b(t)$ can be
taken to be constant in the above equation; see~\cite{jk2,jk3} for a
discussion of the 
validity of this approximation.  Thus, maximizing $\F$ is equivalent to maximizing
$|\tilde{X}(f)|^2$ where $\tilde{X}(f)$ is the \emph{demodulated
Fourier transform} (DeFT) defined as
\be \label{eq:DeFT}
\tilde{X}(f) = \int x(t)e^{-i\Phi(t;f,\vec{\lambda}_d)}\,dt \,.
\ee
With this approximation, the master equation is obtained by solving
\be \label{eq:masterapprox}
\left.\frac{\partial |\tilde{X}(f,\vec{\lambda}_d;
f_{(0)},\vec{\lambda})|^2}{\partial
f}\right|_{\vec{\lambda}_d,f_{(0)},\vec{\lambda}} = 0\,.
\ee
The details of the calculation are given in appendix
\ref{sec:appendix}. The result is:  
\be \label{eq:masterdemod}
f(t) - F_0(t) = \vec{\zeta}(t)\cdot ({\mathbf{n}}-{\mathbf{n}}_d)
\ee
where
\be \label{eq:F0def}
F_0(t) =f_{(0)} + \sum_{k=1}^{s}{\Delta f_{(k)} \over k!}\left(\Delta
t\right)^k \, ,
\ee
and
\ba \label{eq:zeta}
\vec{\zeta}(t) &=& \left(F_0(t) + \sum_{k=1}^{s}{f_{d(k)} \over k!}
\left(\Delta t\right)^k\right)\frac{\mathbf{v}(t)}{c} \\
&+&
\left(\sum_{k=1}^{s} {f_{d(k)} \over (k-1)!}
\left(\Delta t\right)^{k-1}\right)\frac{\mathbf{r}(t) -\mathbf{r}(t_0)}{c} \, .
\nonumber
\ea
The $\Delta f_{(k)}$'s are the residual spin-down parameters: $\Delta
f_{(k)}=f_{(k)} - f_{d(k)}$, and $\mathbf{r}(t)$ is the position of
the detector in the SSB frame. As expected, if $\Delta\vec{\lambda}=0$
so that $\mathbf{n}=\mathbf{n}_d$ and $f_{(k)}=f_{d(k)}$, then
$f(t)=f_{(0)}$.  Furthermore, it is clear that this master equation is
\emph{qualitatively} similar to equation (\ref{eq:master1}) except for a
constant frequency offset $\vec{\zeta}\cdot \mathbf{n}_d$.  Thus, many
of the methods obtained for the non-demodulated case will still be
valid.

\subsection{Implementation details}
\label{subsec:implementdemod}

As mentioned above, the master equations (\ref{eq:master1}) and
(\ref{eq:masterdemod}) are qualitatively similar except for a
constant frequency offset.  Thus, many of the earlier results are
still valid with some minor modifications which we now explain.  

\noindent\textbf{\emph{Resolution in parameter space:}} 
The formula for the resolution in $f_k$ space is the same as given
in equation (\ref{eq:deltafk}).  However, since we can make
$\Tcoh$ much larger than before, the resolution can be made much more
finer.  Thus, for the first spin-down parameter, instead of equation
(\ref{eq:deltaf1}), we would have 
\be
\delta f_{(1)} = (3.7\times 10^{-13}\textrm{Hz/s})\times
\frac{365\textrm{days}}{\Tobs} \cdot\frac{1\textrm{day}}{\Tcoh}\,. 
\ee
Furthermore, the restriction due to the length of $\Tcoh$ (see equation
(\ref{eq:taunondemod})) is no longer an issue. 
  
As for the sky-positions, using the approximation given in equation
(\ref{eq:xiapprox}), the estimate of the resolution in the sky
proceeds in the same way as the derivation of equation
(\ref{eq:annulisize}). The results of equations
(\ref{eq:minannulus}) and (\ref{eq:maxannulus}) are still valid, the
only change being that $\Tcoh$ is now of the order of a day.  Therefore,
we rewrite equations (\ref{eq:minannulus}) and (\ref{eq:maxannulus})
as:
\ba &&\left(\delta\phi\right)_{\textrm{\mbox{\tiny{min}}}} 
= \frac{c}{v}\frac{\delta
  f}{\hat{f}} = \frac{c}{v\hat{f}\Tcoh}  \\ && =  1.0\times
  10^{-3}\,\textrm{rad}\times \left(\frac{1\textrm{day}}{\Tcoh}\right)
\left(\frac{500\,\textrm{Hz}}{\hat{f}}\right) \left(\frac{10^{-4}}{v/c}\right)\nonumber
\label{eq:minannulusdemod} \ea
and 
\ba
&& \left(\delta\phi\right)_{\textrm{\mbox{\tiny{max}}}} =
\sqrt{\frac{\delta f}{\hat{f}}\frac{c}{v}} = \sqrt{\frac{c}{v\hat{f}\Tcoh}}
\\ && = 1.5\times
10^{-2}\,\textrm{rad} \times 
\left(\frac{1\textrm{day}}{\Tcoh}\right)^{\frac{1}{2}} \left(\frac{500 
\textrm{Hz}}{\hat{f}}\right)^{\frac{1}{2}}\left(\frac{10^{-4}}{v/c}\right)^{\frac{1}{2}}
\,. \nonumber \label{eq:maxannulusdemod}
\ea
The sky resolution obtained from equation (\ref{eq:minannulusdemod})
is therefore about 5 times better than in the non-demodulated case
obtained from equation (\ref{eq:minannulus}).\\

\noindent\textbf{\emph{Sky-patch size:}} Unlike in the non-demodulated case,
since we are removing the frequency modulation of the signal
beforehand, there is now, except for computational constraints, 
no restriction at all on the coherent
integration time $\Tcoh$.  Typically, $\Tcoh$
will be taken to be of the order of a day.  However, the price
we pay for this is that the demodulation is not valid for arbitrarily
large patches. The {\it patch size} is determined by the largest 
fractional loss of sensitivity (e.g., the $\F$ value) we are willing to 
tolerate from a true signal with certain mismatch parameters 
$\Delta\vec{\xi}$.

If we have demodulated for a direction
${\mathbf{n}}_d$ in the sky, how different can ${\mathbf{n}}$ be
from ${\mathbf{n}}_d$ so that the loss in the signal power does not
become unacceptably large?  In order to answer this question, we would
have to analyze the parameter space metric defined in terms of the
mismatch~\cite{owen}.  The analysis of the metric will be presented
elsewhere, but in this paper we shall just use a conservative estimate
for the size of the sky-patch.  
 
To estimate the size of the sky-patch, first note that the quantity
$\vec{\mathbf{\zeta}}$ appearing in equation (\ref{eq:masterdemod})
is, to a very good approximation, given by
\be 
\label{eq:xiapprox} 
\vec{\mathbf{\zeta}}(t) \approx \hat{f}(t)\cdot\frac{\mathbf{v}(t)}{c} 
\ee
where, as before, $\mathbf{v}$ is the velocity of the detector in the
SSB frame.  The velocity $\mathbf{v}(t)$ is the sum of two components,
the velocity $\mathbf{v}_y$ due to the yearly motion around the sun
and the velocity $\mathbf{v}_d$ due to the rotation of
Earth around its axis: $\mathbf{v} = \mathbf{v}_y+\mathbf{v}_d$.
For reference, for the GEO detector, the magnitude $v_y$ is
about $10^2$ times larger than $v_d$.  The estimate of the sky-patch
size proceeds roughly like the estimate of the pixel size in equation
(\ref{eq:minannulusdemod}) except for one difference.  If we take the coherent
integration time $\Tcoh$ to be roughly of the order of less than a
day, say a third of a day, then the relevant velocity is
$\mathbf{v}_d$.  Thus, the sky-patch size $h$ is roughly given by
\be
h \approx \frac{c}{v_d}\frac{\delta f}{\hat{f}} = \frac{c}{v_d \hat{f}\Tcoh}\,.
\ee
Since $v_d$ is roughly 100 times smaller than $v$, $h\approx 100
(\delta\phi)_\min$.  Thus, a typical sky-patch consists of about 100
pixels on a side.  It should be emphasized that this is only an
educated guess and is not likely to be valid for larger $\Tcoh$. \\

\noindent\textbf{\emph{Validity of the look up tables:}} Again using
the approximation given 
in equation (\ref{eq:xiapprox}), the number of frequency bins for
which the LUT is valid can be estimated in a similar way as in the
non-demodulated case.  The master equation is 
\be \Delta f := f-F_0 = \hat{f}\frac{\mathbf{v}}{c}\cdot
({\mathbf{n}}- {\mathbf{n}}_d) \ee
Rewrite the equation as
\be \frac{1}{\hat{f}} = \frac{1}{\Delta f}\frac{v}{c}\cos\phi -
\frac{1}{\Delta f}\frac{\mathbf{v}\cdot{\mathbf{n}}_d}{c} \, .\ee
Keeping $\Delta f$ fixed and differentiating with respect to $\hat{f}$ leads to
\be \frac{d \hat{f}}{\hat{f}^2} =  \frac{1}{\Delta
f}\frac{v}{c}\sin\phi\,d\phi \ee
As before, define $\kappa$ and $r$ by $d\hat{f} = \kappa\delta f$ and
$d\phi = r\left(\delta\phi\right)_{\min}$.

Substituting these definitions in the above equation yields
\be \label{eq:nfreqvalid0} \kappa = \frac{rf_0}{\Delta f}\sin\phi \ee 
There are now two cases to look at, namely when $\phi$ is close to
$\pi/2$ or when it is close to $0$ (or $\pi$). First the easy case
when $\phi \sim \pi/2$.  Here the width of the annuli is roughly the
same as the pixel size: 
$\delta\phi \sim \left(\delta\phi\right)_\min$.
Thus, if $h$ is the length of a side of the sky-patch (assumed to be
square) then the number of annuli in the sky patch is $h/\delta\phi$
which means $\Delta f = \delta f \cdot (h/\delta\phi)$. Substituting
this in equation (\ref{eq:nfreqvalid0}) and also setting $\phi=\pi/2$
finally leads to the result
\be 
\label{eq:nfreqvalid}
\kappa = \kappa_0 :=  \frac{r}{hv/c} \, . 
\ee
Now turn to the large annulus case.  The annulus size is given by
$\delta\phi \sim (\left(\delta\phi\right)_\max)$ 
and again $\Delta f = \delta f \cdot
(h/\delta\phi)$. As for the numerator of equation (\ref{eq:nfreqvalid}),
take the smallest value of $\sin\phi$, i.e. when $\phi$ is no bigger
than a pixel so that $\sin\phi \sim
\left(\delta\phi\right)_{\min}$.  Substituting these estimates
leads to 
\be \kappa =\left( \frac{r}{hv/c}\right) \sqrt{\frac{\delta f}{f_0}\frac{c}{v}} =
\kappa_0\left(\delta\phi\right)_{\max} \,.\ee
From equations (\ref{eq:maxannulusdemod}) and (\ref{eq:nfreqvalid}) 
we see that typically, $\kappa$ for
the thick annulus case is about 100 times smaller than for the
thin annulus case.

\subsection{Statistics}
\label{subsec:demodstatistics}

This subsection describes the statistics of the Hough map, 
the $\F$-statistic, the optimal thresholds and the sensitivity.  The
discussion closely parallels that of section \ref{sec:statistics};
here we simply point out some of
the differences that arise when the $\F$-statistic is considered
instead of the normalized power.  

Just as the distribution of the normalized $\rho_k$ power in subsection
\ref{subsec:statistics} turned out
to be related to the $\chi^2$ distribution with 2 degrees of freedom,
one might intuitively expect that the distribution of $\F$ should also
be related to a $\chi^2$ distribution.
However, since $\F$ is constructed from the four filters given in 
equation (\ref{eq:quadratures}), it turns out that the distribution of
$2\F$ is a non-central $\chi^2$ distribution with \emph{four} degrees
of freedom.  As before, we shall denote the non-centrality parameter
by $\lambda$, and it turns out to be
\be \label{eq:noncentralF}
\lambda = (h|h)
\ee
where the inner product $(\cdot |\cdot )$ has been defined in equation
(\ref{eq:innerprod}).  
A word of caution: while we use the same symbol for the non-centrality
parameter as in the non-demodulated case, this definition is different from
that of equation (\ref{eq:noncentral}).  

Thus, the distribution of $\F$ is
\ba
p(\F|\lambda) &=& 2\chi^2(2\F|4,\lambda) \\
&=&\left(\frac{2\F}{\lambda}\right)^{1/2}I_1(\sqrt{2\F\lambda})\,
\exp{\left(-\F-\frac{\lambda}{2}    
\right)} \nonumber 
\ea
where $I_1$ is the modified Bessel function of the first order.  In
the absence of a signal, this reduces to
\be
p(\F |\,0) = \F e^{-\F} \,. 
\ee
We select frequency bins by setting a threshold $\F_\th$ on the value
of the $\F$-statistic in that frequency bin. Given $\F_\th$,
the probabilities for false alarm, false detection and detection are
defined analogous to equations (\ref{eq:probeta}), (\ref{eq:powerfa})
and (\ref{eq:powerfd}):
\begin{eqnarray}
\alpha(\F_\th) &=& \int_{\F_\th}^{\infty} p(\F|\,0)\,d\F \nonumber \\ &=&
(1+\F_\th)e^{-\F_\th}\, ,
\label{eq:fstatfa} \\
\beta(\F_\th|\lambda) &=& \int_0^{\F_\th}p(\F|\lambda)\,d\F\,,
\label{eq:fstatfd} \\
\eta(\F_\th|\lambda) &=& \int_{\F_\th}^{\infty} p(\F|\lambda)\,d\F\,.
\label{eq:probetaF}
\end{eqnarray}
The relation between $\alpha$ and $\F_\th$ is different from
the relation $\alpha=e^{-\rho_\th}$ in the non-demodulated case.  
For small signals, $\eta(\F_\th|\lambda)$ can be expanded as
\be \label{eq:etaapproxF}
\eta = \alpha + \frac{\lambda\F_\th^2}{4}e^{-\F_\th}+ \mathcal{O}(\lambda^2)\,.
\ee
Once
again we will approximate this distribution by a binomial.  In fact,
we expect the binomial approximation to be better in this as compared
to the non-demodulated search because, typically,
$\Tcoh$ will now be larger and thus the signal will see a more
representative `average' of the detector antenna pattern.  Finally,
the expressions for the false alarm
and false dismissal probabilities in the Hough plane are the same as
in equations (\ref{eq:houghalpha}) and (\ref{eq:houghbeta}) but again with
the $\lambda$'s and $\eta$'s as above.  

\begin{figure}
    \includegraphics[height=6cm]{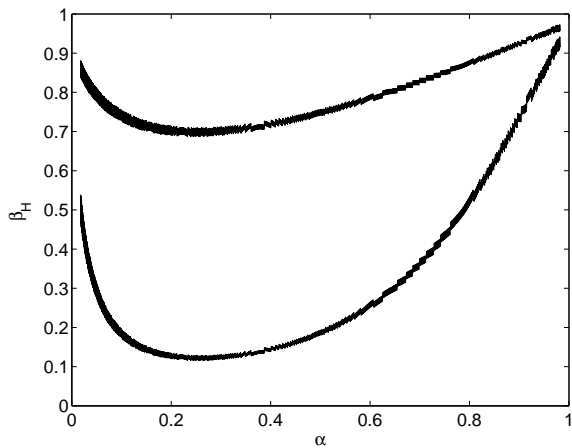}
    \caption{\small{Graph of $\beta_H$ as a function of
    $\alpha=(1+\F_\th)e^{-\F_\th}$ for $\lambda=0.2$ (upper curve) and
    $\lambda=0.4$ (lower curve). Both curves correspond to
    $\alpha_H=0.01$ and $N=1000$.  The minimum of $\beta_H$ occurs
    roughly at $\alpha=0.26$ which corresponds to $\F_\th=2.6$. }}\label{fig:minbetaF}  
\end{figure}

\begin{figure}
    \includegraphics[height=6cm]{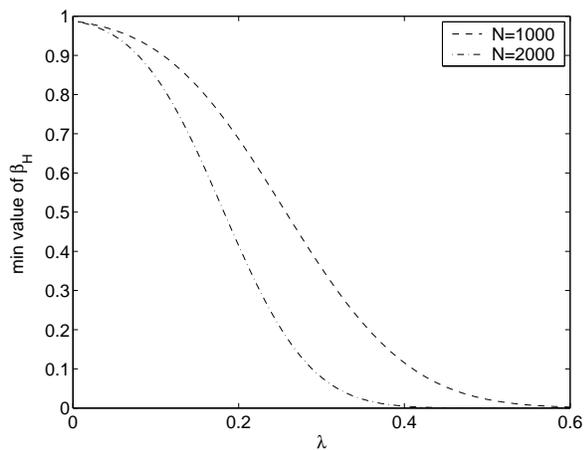}
    \caption{\small{Graph of the minimum of $\beta_H$ as a function of
    $\lambda$ for $N=1000$ and $2000$.  Both curves correspond to
    $\alpha_H=0.01$.}}\label{fig:minvalbetaF}  
\end{figure}

With the above definitions at hand, we are now ready to optimize the
thresholds $\F_\th$ and $n_\th$ using the procedure described in
subsection \ref{subsec:optimalthr}.  The differences from that subsection
are simply in the dependence of $\alpha$ on $\F_\th$ and of $\eta$ on
$\alpha$.  The solution for $n_\th$ obtained by inverting the equation
$\alpha_H(n_\th,\alpha,N)$ given in figure
\ref{fig:nthr} and the analytic approximation of equation
(\ref{eq:nthrapprox}) are unchanged. The graph of $\beta_H$ as a
function of $\alpha$ is however, now different.  The result is shown
in figure \ref{fig:minbetaF}.  The optimal value for the threshold
turns out to be $\F_\th^\star=2.6$ corresponding to a false alarm rate of
$\alpha^\star=0.26$. The minimum value of $\beta_H$ achieved by these
thresholds is plotted in figure \ref{fig:minvalbetaF} as a function of
$\lambda$.  

Finally, let us calculate the sensitivity of the search and obtain the
analog of equation (\ref{eq:sensitivity}).  The starting point is
again equation (\ref{eq:solvebeta}) but now $\eta$ is related to
$\alpha$ by equation (\ref{eq:etaapproxF}) and $\alpha$ is related to
$\F_\th$ by equation (\ref{eq:fstatfa}).  Then, ignoring terms of
$\mathcal{O}(\lambda^2)$ we get the linear approximation for
$\beta_H$:
\be \label{eq:solvebetaapproxF}
\beta_H = \frac{1}{2}\textrm{erfc}\left(-\textrm{erfc}^{-1}(2\alpha_{H}^{\star})
+ \frac{1}{4}\Theta e^{-\F_\th^\star}(\F_\th^\star)^2\lambda\right)
\ee
where, as before, $\Theta$ is given by equation (\ref{eq:c}).  Solving for
$\lambda$ in the large $N$ limit leads to
\be
\lambda \approx \frac{4\mathcal{S}}{(\F_\th^\star)^2
e^{-\F_\th^\star}}\sqrt{\frac{2\alpha^\star(1-\alpha^\star)}{N}} \approx  \frac{12.73}{\sqrt{N}}
\ee
where, to obtain numerical values, we have taken $\alpha_{H}^{\star}=0.01$,
$\beta_H^{\star}=0.1$.
Now average over the parameters $(\iota,\psi,\alpha,\delta)$ and obtain 
\be \label{eq:sensitivityF}
h_0 = \frac{8.92}{N^{1/4}}\sqrt{\frac{S_n}{\Tcoh}} = 8.92\,
N^{1/4}\sqrt{\frac{S_n}{\Tobs}} \,.
\ee
In the second step we have assumed $\Tobs=N\Tcoh$ which is valid only
if there are no gaps in the data.

As expected, equation (\ref{eq:sensitivityF}) is identical to equation 
(\ref{eq:sensitivity}) except for a slightly different numerical
factor.  Thus for comparable values of $\Tcoh$ and $N$, the two
versions of the Hough transform search are very similar in
sensitivity but the search with demodulated data does not have
any restriction on $\Tcoh$ and will thus lead to a much greater
sensitivity, though over a smaller region in parameter space.  
Thus, if we estimate the astrophysical range of the search as in
equation (\ref{eq:range}), we obtain:
\ba
d &=& 15.4\,\textrm{kpc}\times \left(
\frac{N}{365}\right)^{\frac{1}{4}} \left(
\frac{I_{zz}}{10^{38}\textrm{kg-m$^2$}}\right)
\left(\frac{f_r}{500\textrm{Hz}} \right)^2  \nonumber \\
&&\times\left( \frac{\epsilon}{10^{-6}}\right)
\left(\frac{\Tcoh}{1\textrm{day}}\right)^{\frac{1}{2}}\left(
\frac{10^{-46}\textrm{Hz}^{-1}}{S_n}\right)^{\frac{1}{2}}  \,.
\ea
Here we have taken a coherent integration time of $1$day and a total 
observation time of $1$yr as the reference values.  

\section{Conclusions}
\label{sec:conclusion}

Let us summarize the main ideas and results presented in this
work.  Since it is not feasible to perform large parameter space
searches using the matched filter with presently available computing
power, we began by emphasizing the need for hierarchical searches
demonstrating the need for an incoherent and computationally inexpensive
search method.  The Hough transform is an example of such a method.
It looks for patterns in the frequency time plane by constructing a
histogram in parameter space based on the consistency of observations
in the time-frequency plane with an underlying model describing the
pattern. We have given a general description of the Hough transform
and shown its relevance for pulsar searches.  
 
We have presented two versions of the Hough transform search. The
first version takes simple Fourier
transforms as input data.  This restricts the time baseline of the different
segments but it allows us to search over a large sky-patch.  The
second version takes input data which has been demodulated to
remove the effects of Earth's motion and the spin-down of the star;
this is achieved by using the $\F$-statistic.  We have presented some
technical details for both flavors of the search.  In particular, we
show how to solve the master equation in the two cases and how the use
of look-up tables can lead to a large saving in computational cost.   

We have also
analyzed the statistics for both cases and we saw that we need to
choose two thresholds: the threshold $\rho_\th$ or $\F_\th$ on the
coherent statistic used in the two cases, and the threshold $n_\th$ on
the number count in the Hough maps.  These thresholds have been chosen
in such a way that we get the lowest possible false dismissal rate for
a given choice of the false alarm rate.  We also estimate the
sensitivity of the two flavors of the 
Hough transform and we find that for the same
value of $\Tcoh$ and $N$, both variations have comparable sensitivity,
which improves as $N^{-1/4}\Tcoh^{-1/2}$, as would be expected for 
an incoherent method that builds on coherent sub-steps. When compared to 
the sensitivity that a fully coherent search in a very large parameter 
space would have for the same total observation time $\Tobs$, the 
Hough methods are worse by roughly a factor of $N^{1/4}$.  Considering that
the Hough transform can be expected to run very much faster 
than any coherent method, it should therefore be able to survey much 
larger volumes of space than coherent methods, despite its poorer 
sensitivity in any single direction.  This is therefore a potentially
very important method for conducting large-scale gravitational wave
pulsar surveys.

\acknowledgments 

We would like to thank Greg Mendell for many useful discussions and 
for his suggestions after reading one of the last drafts of this paper.
We also acknowledge useful discussions with Carsten Aulbert, Curt Cutler,
Steffen Grunewald, Yousuke Itoh, Federico Massaioli, Reinhard Prix,
Linqing Wen, and Lucia Zanello.  We also acknowledge the support of
the Max Planck Society (Albert Einstein Institut) and the Spanish {\it 
Ministerio de Ciencia y Tecnolog\'{\i}a} DGICYT Research Project
BFM2001-0988.

\begin{appendix}

\section{The master equation for the search with demodulated data}
\label{sec:appendix}

Here we detail the steps leading from equation (\ref{eq:masterapprox})
to the master equation (\ref{eq:masterdemod}).

Ignoring the  amplitude modulation and a possible constant phase,
the signal from a pulsar with parameters 
$(f_{(0)}; \vec{\lambda})=(f_{(0)}; {\mathbf{n}},\{f_{(n)}\})$ 
would  be:
\be
h(t,f_{(0)}, \vec\lambda) = e^{i\Phi(t,f_{(0)},\vec\lambda)}  \, ,
 \label{eq:simplesignal}
\ee
where
\be
\Phi(t;f_{(0)},\vec\lambda) =
  2\pi  \left[ f_{(0)} \Delta t_{\bf n}  +
  \sum_{k=1}^s {f_{(k)} \over (k+1)!} \left( \Delta t_{\bf
    n}\right)^{k+1} \right]  
  \, ,
\ee
and
\be
\Delta t_{\bf n}=t_\SSB(t,{\bf n})-t_\SSB(t_0, {\bf n}) \, .
\ee
Here $t_0$ is the time in the detector frame to which the frequency
and spin-down parameters refer to and $t_\SSB$ is time in the SSB
frame.  Neglecting higher order relativistic effects, the detector
time $t$ is related to $t_\SSB$ by
\be t_\SSB (t,{\mathbf{n}}) = t +
\frac{\mathbf{r}(t)\cdot\mathbf{{n}}}{c} \ee 
where $\mathbf{r}(t)$ is the detector position in the SSB frame.

The DeFT of the pulsar signal (\ref{eq:simplesignal}) with respect to the
demodulation parameters $(f,\vec\lambda_d)$ is:
\be \label{eq:deftdef}
\tilde X(f) = \int_{-\frac{1}{2}\Tcoh}^{\frac{1}{2}\Tcoh}
e^{ i[\Phi(t, f_{(0)},\vec \lambda)
-\Phi(t, f,\vec\lambda_d)]
} dt \,. 
\ee
Without any loss of generality, we have taken the coherent time
interval to be centered around $t=0$ so that the integral is from
$-\Tcoh/2$ to $\Tcoh/2$.  

Our goal is  to determine an analytical expression for the value
of $f$ that maximizes the power $P(f)=|\tilde X(f)|^2$ in
terms of  $f_{(0)}$, $\vec\lambda_d$ and $\Delta\vec\lambda$.  To 
do this, first expand $\Delta\Phi(t):=\Phi(t,f_{(0)},\vec{\lambda}) -
\Phi(t,f,\vec{\lambda}_d)$ in powers of $\Delta f_{(k)}:=f_{(k)} -
f_{d(k)}$  and $\Delta \mathbf{n} := \mathbf{n} -
\mathbf{n}_d$, keeping only terms up to linear order: 
\ba
&&\frac{\Delta \Phi(t)}{2\pi} 
\approx \left(f_{(0)}-f\right) {\Delta t_{\nd}}
+ \sum_{k=1}^{s} {\Delta f_{(k)} \over (k+1)!} \left({\Delta
t_{\nd}}\right)^{k+1} \nonumber\\
&&  + \left[ f_{(0)} + \sum_{k=1}^{s} \frac{f_{(k)}}{k!}
\left(\Delta t_{\nd}\right)^k \right]  
\frac{\Delta\mathbf{r}}{c}\cdot \Delta\mathbf{n} \label{eq:expanddeltaphi}
\ea
where $\Delta\mathbf{r}=\mathbf{r}(t)-\mathbf{r}(t_0)$.

Now Taylor expand $\Delta\Phi(t)$ about a fiducial time $t_1$ 
in the interval $-\Tcoh/2\le t_1 \le \Tcoh/2$ again retaining terms
only up to linear order \footnote{ The choice of $t_1$ within
  $(-\Tcoh/2,\Tcoh/2)$ does not 
matter because, the sky-patch in which the demodulation is
valid, is such that the power concentrates in a single frequency bin.}
\ba
\tilde X(f)& \approx & \int_{-\frac{1}{2}\Tcoh}^{\frac{1}{2}\Tcoh}
e^{i\left( \Delta\Phi (t_1) + (t-t_1)
{\displaystyle \partial \Delta\Phi\over \displaystyle\partial t}\right)}dt 
\nonumber\\
& =&e^{i \Delta\Phi (t_1)}
\int_{-\frac{1}{2}\Tcoh}^{\frac{1}{2}\Tcoh}e^{i (t-t_1)
{\displaystyle\partial \Delta\Phi\over \displaystyle\partial t}} dt \ .
\ea
$P(f)$ does not depend on $ \Delta\Phi (t_1)$, and
its maximum is reached for the  value of $f$ that satisfies 
\be
\left. \frac{\partial \Delta\Phi}{\partial t}\right|_{t=t_1} = 0 \,.
\ee
Differentiating (\ref{eq:expanddeltaphi}) with respect to $t$, we get
\ba
& & {\frac{1}{2\pi}}\left.\frac{\partial\Delta\Phi}{\partial
t}\right|_{t=t_1} \\
& & \approx\left( f_{(0)} -f +  \sum_{k=1}^{s} {\Delta f_{(k)}\over k!}
\left(\Delta t_1\right)^k \right)  
\left(1+ {{\mathbf{v}}(t_1)\over c}\cdot {\mathbf{n}}_d \right) \nonumber \\
& & +\left( f_{(0)} +\sum_{k=1}^{s} \frac{f_{(k)}}{k!}
\left(\Delta t_1\right)^k  \right) 
{{\mathbf{v}}(t_1)\over c}\cdot \Delta\mathbf{n} \nonumber \\
& & +\left( \sum_{k=1}^{s}  \frac{f_{(k)}}{(k-1)!}\left(\Delta
t_1\right)^{k-1}  \right) \left(1+ {{\mathbf{v}}(t_1)\over c}\cdot
{\mathbf{n}}_d \right)
\frac{\Delta\mathbf{r}_1}{c}\cdot\Delta\mathbf{n} \nonumber 
\ea
where $\Delta t_1= t_\SSB(t_1,\nd)-t_\SSB(t_0,\nd)$ and
$\Delta\mathbf{r}_1=\mathbf{r}(t_1) -\mathbf{r}(t_0)$.  Setting the
right hand side of this equation to zero and dropping higher order
terms leads to
\ba
& &f - F_0  = \left( F_0+ 
\sum_{k=1}^{s} {f_{d(k)}\over k!} \left(\Delta t_1\right)^k \right) 
{{\mathbf{v}}(t_1)\over c}\cdot \Delta\mathbf{n}  \nonumber \\
& & + \left( \sum_{k=1}^{s} {{f_{d(k)}} \over (k-1)!}
 \left(\Delta t_1\right)^{k-1}  \right)
 \frac{\Delta\mathbf{r}_1}{c}\cdot \Delta\mathbf{n}  
\ea
where $F_0$ is as given in equation (\ref{eq:F0def}).
All the dependence on the residual 
spin-down parameters appears only in the definition of $F_0$ and,
after replacing the arbitrary time $t_1$ by $t$, we
get equation (\ref{eq:masterdemod}).

\end{appendix}
\newpage 


\end{document}